\documentclass[aps,prd,twocolumn,superscriptaddress,nofootinbib]{revtex4-2}
\usepackage{graphicx}
\usepackage{breqn}
\usepackage{mathtools}
\usepackage{amsmath}
\usepackage[dvipsnames]{xcolor}
\usepackage{xcolor}

\usepackage{aas_macros}

\usepackage{mathrsfs}
\usepackage{subfig}
\usepackage{multirow}
\usepackage{stackengine}
\usepackage{makecell}

\usepackage{tablefootnote} 
\usepackage{tabularx}
\usepackage{array} 

\newcommand{\Blos}{B_{p_{\hatbf{n}}p_{\hatbf{n}}\delta}}

\newcommand{\Bperp}{B_{p_{\perp}p_{\perp}\delta}}

\newcommand{\CTTg}{C_{\ell}^{\mathrm{kSZ}^{2}\times\delta_{g}}}
 
\newcommand{\mnus}{\Sigma m_{\nu}}
\newcommand{\hatbf}[1]{\mathbf{\hat{#1}}}
\newcommand{\BTTg}{B^{\mathrm{kSZ},\mathrm{kSZ},\delta_g}}

\newcommand{\Rev}[1]{{{#1}}}
\newcommand{\Corr}[1]{{{#1}}}

\graphicspath{{./}{figures/}}

\begin{document}
\title{A Novel Bispectrum Estimator of the Kinematic Sunyaev-Zel'dovich Effect using Projected Fields} 
\author{Raagini Patki}
\email[Corresponding author:\ ]{rp585@cornell.edu}
\affiliation{Department of Astronomy\char`,{} Cornell University\char`,{} Ithaca\char`,{} NY 14853\char`,{} USA.}
\author{Nicholas Battaglia}
\affiliation{Department of Astronomy\char`,{} Cornell University\char`,{} Ithaca\char`,{} NY 14853\char`,{} USA.}
\affiliation{Université Paris Cité\char`,{} CNRS\char`,{} Astroparticule et Cosmologie\char`,{} F-75013 Paris\char`,{} France} 
\author{J.~Colin Hill}
\affiliation{Department of Physics\char`,{} Columbia University\char`,{} New York\char`,{} NY 10027\char`,{} USA.}

\begin{abstract}
With the advent of current and future high-resolution CMB experiments, the kinematic Sunyaev-Zel’dovich (kSZ) effect has become a unique observational probe of the distribution of baryons and velocities in the Universe. In this work, we propose a novel binned bispectrum of the form temperature-temperature-density to extract the late-time kSZ effect from cleaned CMB maps. Unlike `kSZ tomography' methods, this estimator can use any tracer of the large-scale structure density field projected along the line-of-sight and does not require individual redshifts. 
With our method, we forecast signal-to-noise ratios (SNR) of $\sim$100-200 for the upcoming Simons Observatory (SO) and CMB-S4 correlated with a galaxy sample from WISE that is restricted to the linear regime. 
We also extend galaxy modes into the non-linear regime and explore this harmonic space to show that the SNR peaks for squeezed triangles that have a short (linear) density mode and long temperature modes in harmonic space. The existing kSZ$^{2}-$density projected-fields estimator compresses the rich information contained in this bispectrum across various scales and triangle shapes. Moreover, we find that the lensing correction to this estimator's signal is relatively small for high-SNR squeezed triangle configurations. We study the dependence of this kSZ signal on $\Lambda$CDM parameters for SO and CMB-S4 and forecast initial constraints on the sum of neutrino masses while restricting to the linear galaxy bias regime. Our work illustrates the potential of the projected-fields kSZ bispectrum as a novel probe of baryonic abundance and beyond-$\Lambda$CDM cosmology with upcoming precision measurements. 
\end{abstract}

\maketitle

\section{\label{sec:intro}Introduction}
In recent years, high-resolution observations of the Cosmic Microwave Background (CMB) by ground-based telescopes such as the Atacama Cosmology Telescope (ACT) \cite{ACT2016} and the South Pole Telescope (SPT) \cite{SPT2011} have made secondary anisotropies detectable at high significance. 
These observations open a window into the late-time Universe, going beyond the primary anisotropies that originated in the early Universe, which were measured by all-sky satellites such as \textit{Planck} \cite{Planck2018}  
The kinematic Sunyaev-Zel’dovich (kSZ) effect \cite{SZ1972, SZ1980, Ostriker1986} arises due to Compton-scattering between CMB photons and free electrons with a non-zero bulk velocity; the resulting Doppler boosting of photons leads to a secondary CMB anisotropy that is proportional to the electron momentum along the line-of-sight (LOS). Thus, the kSZ effect has emerged as a probe of both astrophysics (e.g.\,\cite{F16, Battaglia2017, BattagliaHill2019, Amodeo2020, Bolliet2022}) and cosmology (e.g.\,\cite{Bhattacharya2008, Mueller2014DE, P23, Mueller2014, Tishue25_mnu, Munch}), through its dependence on baryon density and peculiar velocity respectively.   

Since the kSZ effect maintains the blackbody spectrum of the primary CMB, it cannot be isolated from observed CMB maps using component separation techniques alone. The kSZ signal sourced by electrons in the late-time Universe is then detected by combining CMB data with Large-Scale Structure (LSS) observations through various statistical methods or `estimators'. The pairwise momenta of galaxies (e.g.\,\cite{Ferreira1999}) was applied to ACT data along with BOSS spectroscopic galaxies leading to the first detection of the kSZ effect in 2012 \cite{Hand2012}, followed by others in \textit{Planck}, SPT, and ACT data \cite{Planckpairwise, SPTpairwise, Calafut2021} with this approach. The velocity-weighted stacking method \cite{Ho2009, Shao} has been used to constrain the baryon density profile in ACT data with BOSS galaxies \cite{Schaan2021} and with DESI photometric LRGs \cite{Hadzhiyska2024}. On the other hand, velocity reconstruction (e.g.\,\cite{Deutsch2018, Smith2018}) has been forecasted to probe primordial non-Gaussianity \cite{Munch, Kumar2022} and the Universe on the largest possible scales \cite{giri2020, Cayuso2023}; the first detection with this method was done recently in ACT data with DESI photometric LRGs \cite{FMcCarthy2024}. 

All of these `kSZ tomography' estimators were shown to be mathematically equivalent - different ways of representing the $\langle Tgg \rangle$ statistic  - a bispectrum\footnote{A bispectrum refers to the harmonic space equivalent of a three-point correlation function. This is analogous to a power spectrum in harmonic space being the Fourier transform of a two-point correlation function.} between two powers of a galaxy field and one power of CMB temperature \cite{Smith2018}. On the other hand, \Rev{in this work, we study a distinct, `full' bispectrum of the form $\langle TTg \rangle$ for the first time}. An existing method, the so-called `kSZ$^{2}\times\delta_g$' or projected-fields power spectrum involves filtering and squaring a CMB map in real space before cross-correlating with a projected-LSS field \cite{Dore2004, D05}. \Rev{This kSZ$^{2}$ method sums the $\langle TTg \rangle$ statistic across all triangle shapes, and is thus a \textit{compressed} version of the `full' bispectrum estimator that we have newly proposed and explored in this work.} 

The kSZ bispectrum here and the `kSZ$^{2}$' estimator both utilize LSS fields that are \textit{projected} along the LOS, and thus only require a rough redshift distribution of a sample of tracers ($dn/dz$). Considering galaxies as the LSS tracer ($\delta_g$), \Rev{these methods can thus utilize even those large-volume galaxy samples that have large photometric redshift uncertainties, such as the WISE} \cite{WISE} and unWISE \cite{unWISE2019} catalogs and upcoming data from the Rubin Observatory (VRO) \cite{VRO2019}. Moreover, these two kSZ estimators have the distinct advantage of being applicable to \textit{any tracer} of the LSS density field (instead of $\delta_g$), such as galaxy/CMB weak lensing convergence maps \cite{Bolliet2022} and 21-cm fluctuations to probe reionization \cite{LaPlante_2020_21kSZkSZ_BiS}, etc. In contrast, the \Rev{`kSZ tomography' methods need estimates of individual galaxy redshifts to make significant detections in 3D (see e.g.\,\cite{Flender2016, BlochJohnson2024, FMcCarthy2024, Lague2024}), and are limited to either spectroscopic galaxies or photometric samples having small per-object photo-z errors (e.g.\,DESI LRGs} with $\sigma_z/(1+z)\lesssim0.02$ \cite{DESILRG_2023}).

So far, there have been two kSZ$^{2}$$\times\delta_g$ measurements at $\sim$3-5$\sigma$ in \textit{Planck} maps using WISE \cite{Hill2016, F16} and unWISE \cite{Kusiak2021} data. The proposed projected-fields kSZ bispectrum avoids a convolution over the CMB filter that otherwise occurs with the kSZ$^{2}$ method. Moreover, we bin the $\langle TTg \rangle$ statistic in harmonic space, instead of compressing information across all triangle shapes. \Rev{Hence, this novel `full' bispectrum contains richer information that is better separated across different scales.} This also allows us to restrict $\delta_g$ modes to the linear regime to avoid uncertainties in the Halo Occupation Distribution (HOD) and non-linear galaxy bias, while maintaining high-fidelity: we forecast high signal-to-noise ratios of $\sim$100-200 with our estimator for SO$\times$WISE and CMB-S4$\times$WISE, which are comparable or higher than those forecasted with the kSZ$^{2}$ method using unrestricted $\delta_g$ modes \cite{Bolliet2022, P23}.    

We assume the best-fit $\Lambda$CDM model from Planck-2018 \cite{Planck2018} as our fiducial model. This paper is organized as follows: In Section II we describe the theoretical formalism of our proposed kSZ bispectrum and provide details about its binning, the lensing contribution to this signal, and its covariance matrix. We summarize survey specifications, analysis choices, and numerical implementation in Section III. We present forecasts with this estimator for SO$\times$WISE and CMB-S4$\times$WISE in Section IV and discuss its information content across triangle shapes. In Section V, we study the cosmological dependence of this novel kSZ bispectrum on $\Lambda$CDM parameters. We also illustrate its potential beyond measuring the late-time baryon abundance by forecasting constraints on the sum of neutrino masses with this probe. We discuss these results and the future outlook with this novel estimator to conclude in Section VI. 

\section{Theoretical formalism}\label{theory}
We begin with an overview of the kSZ effect and the existing kSZ$^{2}\times\delta_g$ estimator that has been used to detect it using projected-fields of matter overdensity. We then define our novel projected-fields kSZ bispectrum and detail its implementation with a binned bispectrum approach.

The kSZ effect gives rise to a secondary shift in the CMB temperature anisotropies $\Theta^{\mathrm{kSZ}}\left( \hatbf{n}\right) \equiv \Delta T^{\mathrm{kSZ}}/T_{\mathrm{CMB}} \left( \hatbf{n}\right)$ in a direction $\hatbf{n}$ on the sky, given by (c $\equiv 1$): 
\begin{equation}\label{kSZ}
\Theta^{\mathrm{kSZ}}\left( \hatbf{n}\right)= -\int d\eta \hspace{0.05 cm} g(\eta) \hspace{0.1 cm} \mathbf{p_{e}}\cdot\hatbf{n},     
\end{equation}
where $\eta(z)$ is the comoving distance to redshift $z$ and $\mathbf{p_{e}} = (1 + \delta_{e}) \mathbf{v}_{e}$ is the electron momentum field, with $\delta_{e}$ and $\mathbf{v}_{e}$ being the overdensity and peculiar velocity of electrons, respectively. $g(\eta) = e^{-\tau} d\tau/d\eta$ is the visibility function, and $\tau$ is the optical depth to Thomson scattering. $\left(d\tau/d\eta\right) = \sigma_{\mathrm{T}}\, n_e/(1+z)$, where $\sigma_{\mathrm{T}}$ is the Thomson scattering cross-section and $n_e$ is the free electron number density, so that
\begin{equation}
\Theta^{\mathrm{kSZ}}\left( \hatbf{n}\right)= -\sigma_{\mathrm{T}} \int \frac{d\eta}{1+z} e^{-\tau} n_e\left( \hatbf{n}, \eta \right) \mathbf{v_{e}}\left( \hatbf{n}, \eta \right)\cdot\hatbf{n}.     
\end{equation}

In this work, we consider galaxies as a tracer of matter in the LSS. However, the same formalism can be similarly applied to any other tracers of the underlying matter density field, such as quasars, \Rev{21-cm fluctuations (e.g.~\cite{Ma2018, LaPlante_2020_21kSZkSZ_BiS})}, or weak lensing convergence (e.g.~\cite{Bolliet2022}). $\delta_{g}(\hatbf{n})$ is the galaxy overdensity field \textit{projected} along the line-of-sight (LOS), and is given by:
\begin{equation}\label{delg}
    \delta_{g}(\hatbf{n}) = \int_{0}^{\eta_{\max}} d\eta \hspace{0.07 cm} W^{g}(\eta) \hspace{0.07 cm} \delta_{m}(\eta\hatbf{n}, \eta),
\end{equation}
where $\delta_{m}(\eta\hatbf{n}, \eta)$ is the matter (fractional) overdensity field in 3D. $\eta_{\max}$ is the maximum comoving distance of the galaxy sample, and its projection kernel is $W^{g}(\eta) = b_g p_s(\eta)$, where $b_g$ is the galaxy bias and $p_s(\eta) \propto dn/d\eta$ is the statistical redshift distribution of the number of galaxies, normalized to have a unit integral. 

Since the kSZ effect has the same (blackbody) frequency dependence as the primary CMB, it cannot be isolated using component-separation methods alone. Thus, we follow previous kSZ works utilizing projected-fields \cite{F16, Hill2016, Kusiak2021, P23} and use a foreground-cleaned map of CMB temperature anisotropies ($\Theta$), whose total auto-power spectrum $C_{\ell}^{\mathrm{tot}} \approx \left(C_{\ell}^{TT} + C_{\ell}^{\mathrm{kSZ}} + N_\ell\right)$ includes contributions from auto-power spectra of the lensed primary CMB ($C_{\ell}^{TT}$), the kSZ effect ($C_{\ell}^{\mathrm{kSZ}}$), and detector noise and residual foregrounds ($N_\ell$). As explained in Section \ref{survey}, in this work, we follow \cite{F16, P23} and avoid additional contributions from the integrated Sachs-Wolfe (ISW) effect by not considering the large scales where it is found to be significant ($\ell < 100$). 
In practice, telescopes observe the CMB through a finite beam $b(\ell)$ modeled by a Gaussian as: 
\begin{equation} \label{beam}
b(\ell) = \mathrm{exp} \left(-\frac{1}{2}\ell(\ell+1)\frac{\theta_{\mathrm{FWHM}}^{2}}{8\,\mathrm{ln}\,2} \right),    
\end{equation}
where $\theta_{\mathrm{FWHM}}$ is the full width at half maximum in radians. The observed CMB temperature anisotropies are therefore limited by the telescope's resolution and are related to the true CMB anisotropies in harmonic space as $\Theta_b(\mathbf{\ell}) = b(\ell)\Theta(\mathbf{\ell})$. 

\subsection{The Existing Projected-Fields kSZ Power Spectrum $\langle \mathrm{kSZ}^2\times\delta_g \rangle$}\label{Clttg}
Free electrons in the LSS are equally likely to be moving towards or away from the observer along the LOS. \Rev{Given this $\mathbf{v_{e}}\rightarrow-\mathbf{v_{e}}$ parity, under statistical isotropy, the cross-power spectrum $\langle\Theta^{\mathrm{kSZ}}\times\delta_g\rangle$ between a cleaned CMB map and a projected LSS field is statistically zero at our scales of interest ($\ell > 100$), due to cancellations among equally likely positive and negative contributions to the kSZ effect when summed along the LOS} (although the linear-ISW effect can have detectable contributions at $\ell < 100$ \cite{F16}). To avoid this naive cancellation, the so-called `projected-fields' or `kSZ$^{2}\times\delta_g$' estimator \cite{Dore2004, D05} \textit{squares} a cleaned CMB temperature map before cross-correlating with $\delta_g$. This \textit{real} space squaring operation necessitates applying a Wiener filter to the cleaned CMB map to preferentially select angular scales dominated by the kSZ contribution, given by:
$F(\ell) = \left(C_{\ell}^{\mathrm{kSZ}}/C_{\ell}^{\mathrm{tot}}\right)$,
where $C_{\ell}^{\mathrm{kSZ}}$ is the template (theoretical) kSZ power spectrum. The overall filtered CMB map is then given in harmonic space by: 
\begin{equation}
\Theta_f(\mathbf{\ell}) = F(\ell)b(\ell)\Theta(\mathbf{\ell}) \equiv f(\ell)\Theta(\mathbf{\ell}),   
\end{equation}
where $f(\ell) \equiv F(\ell)b(\ell)$. Thus, the existing `projected-fields' estimator is a power spectrum defined as:
\begin{equation}\label{dirac}
\langle \Theta_f^{2}(\mathbf{\ell}) \, \delta_g(\mathbf{\ell^'})\rangle = (2\pi)^{2}\delta_{D}(\mathbf{\ell}+\mathbf{\ell^'}) C_{\ell}^{\mathrm{kSZ}^{2}\times\delta_{g}}.  
\end{equation}
Using the Limber approximation \cite{Kaiser1992, Limber}, this $\mathrm{kSZ}^2$-galaxy cross-power spectrum of projected fields is written as \cite{Dore2004, D05, Hill2016, F16, Kusiak2021, Bolliet2022, P23}:
\begin{equation}\label{Cl-def}
    C_{\ell}^{\mathrm{kSZ}^{2}\times\delta_{g}} = \int_{0}^{\eta_{\max}} \frac{d\eta}{\eta^{2}} W^{g}(\eta) g^{2}(\eta) \mathcal{T}\left(j = \frac{\ell}{\eta}, \eta\right),
\end{equation}
where the `triangle power spectrum' $\mathcal{T}$ is
\begin{equation}\label{triangle}
    \mathcal{T}(j, \eta) =\int \frac{d^{2}\mathbf{q}}{(2\pi)^{2}} f(q\eta) f(|\mathbf{j} + \mathbf{q}|\eta) \, \Blos(\mathbf{q}, -\mathbf{j} - \mathbf{q}, \mathbf{j}).
\end{equation}

The key theoretical statistic here is $\Blos$ - 
the 3-point correlation function (or `bispectrum' in Fourier space) 
between two electron momenta projected along the LOS ($p_{\hatbf{n}} = \mathbf{p}\cdot\hatbf{n}$) and one matter overdensity field. Thus, $\mathcal{T}(j, \eta)$ \textit{compresses} the rich information contained in $\Blos$ by summing over all triangles in a constant-redshift plane (at a distance $\eta(z)$) that have one side of length $j$. Moreover, as 
pointed out in \cite{P23}, the squaring operation results in a convolution over the CMB filter $f(\ell)$, which leads to a mixing of the information content in the signal across different angular scales.  
 
\subsection{The Projected-Fields kSZ Bispectrum $\langle \Theta^{\mathrm{kSZ}}\times\Theta^{\mathrm{kSZ}}\times\delta_g \rangle$}\label{ourest} 
We now propose a novel bispectrum estimator for the kSZ that again utilizes projected-fields of LSS tracers of matter, but circumvents the drawbacks of the existing power spectrum estimator that we have described above. It thus retains the advantage of not requiring accurate inidividual redshifts of the LSS tracers, making it applicable to photometric galaxy surveys (with significant photo-z errors), as well as
galaxy/CMB weak lensing convergence, 21-cm fluctuations, etc. 

Here, instead of squaring the (filtered) CMB map as done earlier \cite{Dore2004, D05, Hill2016, F16, Kusiak2021, Bolliet2022, P23}, we consider a full bispectrum of the form $\langle \Theta^{\mathrm{kSZ}}\times\Theta^{\mathrm{kSZ}}\times\delta_g \rangle$ which is estimated (under statistical isotropy and homogeneity) as:
\begin{equation}\label{diracbi}
\langle \Theta_b(\mathbf{\ell_1}) \, \Theta_b(\mathbf{\ell_2})\,\delta_g(\mathbf{\ell_3})\rangle = (2\pi)^{2}\delta_{D}(\mathbf{\ell_1}+\mathbf{\ell_2}+\mathbf{\ell_3}) B_{\ell_1\ell_2\ell_3}^{\mathrm{kSZ},\mathrm{kSZ},\delta_g},  
\end{equation}
where $\Theta_b \equiv b(\ell)\Theta(\mathbf{\ell})$ and $\delta_g$ are the same as before. Taking a cross-correlation of cleaned CMB maps in Fourier space allows us to avoid the real space squaring operation that leads to a convolution over the filter in the existing $\CTTg$ estimator. This also means that the full bispectrum approach does not require Wiener filtering of the CMB maps, because the LSS is only correlated with the kSZ secondary anisotropies, and not with the primary CMB (in the regime where ISW is negligible). Furthermore, the rich information contained in the $\Blos$ statistic is not compressed across different triangle shapes and retains better scale separation in our bispectrum. 

We define this (`reduced') bispectrum $B_{\ell_1\ell_2\ell_3}^{\mathrm{kSZ},\mathrm{kSZ},\delta_g}$ in the flat-sky limit, since we restrict to small-scales ($\ell > 100$) in both $\Theta_b$ and $\delta_g$ (see Section \ref{analysis}). The bispectrum is thus a function of 2D Fourier modes $\mathbf{\ell_1}, \mathbf{\ell_2}$, and $\mathbf{\ell_3}$ that must form a closed triangle (in an isotropic Universe), as enforced by the Dirac delta function above. Also, since $\Theta^{\mathrm{kSZ}}$ reverses in sign upon reflection along the LOS ($\mathbf{v_{e}}\rightarrow-\mathbf{v_{e}}$), it has `odd' parity. Given that our estimator has two such $\Theta^{\mathrm{kSZ}}$ fields, overall, it is a `parity-even' bispectrum (unlike the parity-odd kSZ bispectrum in \cite{Smith2018}). This implies that it will be statistically non-zero only when the triangle inequality is satisfied ($|\ell_1-\ell_2|\leq\ell_3\leq(\ell_1+\ell_2)$) \textit{and} when $(\ell_1+\ell_2+\ell_3=$ even) \cite{hu_2000_bis, Bucher_2016, CoultonSpergel2019}.    

\begin{figure}
\center
{\includegraphics[width=0.95\columnwidth]{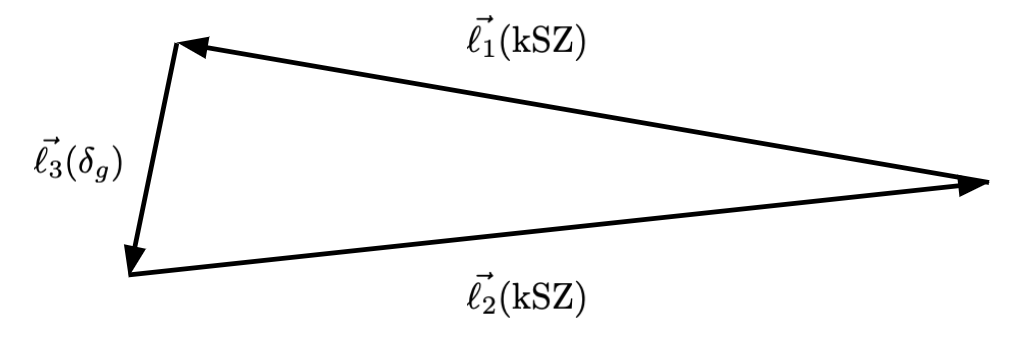} 
}
\caption{Schematic diagram of the type of squeezed triangles that are found to have a significant SNR with the projected-fields $\BTTg$ bispectrum (see Section \ref{fore}). This class of configurations is roughly defined here as consisting of two small-scale (large $\ell>2000$) kSZ modes and one large-scale (small $\ell\lesssim700$) $\delta_g$ mode in harmonic space.}
\label{fig:squeezed}
\end{figure}

From Eqs.\,\eqref{kSZ},\eqref{delg}, and using the Limber approximation for bispectra of projected fields \cite{buch2000}, it follows that:
\begin{dmath} \label{bis} 
B_{\ell_1\ell_2\ell_3}^{\mathrm{kSZ},\mathrm{kSZ},\delta_g} = b(\ell_1)b(\ell_2)\int_{0}^{\eta_{\max}} \frac{d\eta}{\eta^{4}} W^{g}(\eta) g^{2}(\eta) \Blos\left(\frac{\ell_1}{\eta}, \frac{\ell_2}{\eta}, \frac{\ell_3}{\eta}\right),   
\end{dmath}
where $b(\ell)$ is again the beam of the CMB survey. Note that $\Blos$ is also a function of the comoving distance $\eta$, and is the same key theoretical statistic that is involved in both this estimator and the existing $\CTTg$ estimator. Since we are considering different triangle configurations of the $\BTTg$ bispectrum, it becomes even more crucial to adapt the improved model for $\Blos$ \cite{P23} while making theoretical predictions, given that the approximate model is inaccurate for certain triangle shapes. This is particularly important for the class of \textit{squeezed} triangles that have two long sides (i.e. small scales) corresponding to the kSZ modes and one short side (i.e. large scale) corresponding to the LSS tracer in $\ell$-space, such as the one shown schematically in Fig.\,\ref{fig:squeezed}. 
We summarize the improved modeling of $\Blos$ in Appendix \ref{deriv} and refer the reader to \cite{P23} for further details. 
\subsubsection{Binning the Projected-Fields kSZ Bispectrum} 
So far, we have discussed the $\BTTg$ bispectrum for a single triplet of multipoles. However, in practice, the signal-to-noise ratio (SNR) for an individual triplet is expected to be small, making it hard to directly compare future measurements with theory predictions. To address this issue, we use a binned bispectrum approach \cite{bucher_2010, Bucher_2016} that has previously been applied in other contexts such as detecting primordial non-Gaussianity and characterizing foregrounds (e.g. \cite{CoultonSpergel2019, Coulton, Planck2013fNL, Planck2015fNL, rana2018}). Essentially, it bins the $\BTTg_{\ell_1\ell_2\ell_3}$ signal in 3D harmonic space. Given that this bispectrum varies smoothly with its $\ell$ arguments, the binned approach (with an appropriate bin size) would involve very little loss of scale-dependent information. Thus, this approach makes it possible to have statistically significant kSZ detections with future observations using our estimator, without needing to assume a specific template for the expected signal. 

We define the following quantity in terms of Wigner-3j symbols \cite{Bucher_2016, CoultonSpergel2019}:
\begin{equation}
N_{\Delta}^{\ell_1\ell_2\ell_3} = \frac{(2\ell_1+1)(2\ell_2+1)(2\ell_3+1)}{4\pi}\begin{pmatrix}
    \ell_1 & \ell_2 & \ell_3 \\
    0 & 0 & 0
  \end{pmatrix}^2,
\end{equation}
which can be interpreted\footnote{The power spectrum analog is that there are $(2\ell+1)$ possible modes on the celestial sphere for measuring $C_{\ell}$, where maps across the sky are decomposed using spherical harmonics $Y_{\ell m}$.} as the total number of possible triangles on the celestial
sphere with sides $(\ell_1,\ell_2,\ell_3)$. It follows that $N_{\Delta}^{\ell_1\ell_2\ell_3}$ is non-zero only when $\ell_1+\ell_2+\ell_3=$ even. By defining the corresponding all-sky bispectrum as in Eq.\eqref{allsky} (where we follow the convention of \cite{bucher_2010, Bucher_2016, CoultonSpergel2019} instead of the earlier one \cite{hu_2000_bis}), we note that it is related to our flat-sky bispectrum as $\mathscr{B}_{\ell_1\ell_2\ell_3}^{\mathrm{kSZ},\mathrm{kSZ},\delta_g} = N_{\Delta}^{\ell_1\ell_2\ell_3} \BTTg_{\ell_1\ell_2\ell_3}$, for a single multipole-triplet. 

In practice, the binned approach is applied to maps (either from real data or simulations) by dividing the entire multipole range of each of the three fields into subintervals denoted by $\Delta_{a}\equiv[\ell_a,\ell_{a+1}-1]$ where $a = 1,2,...N_{\mathrm{bins}}$ and $N_{\mathrm{bins}}$ is the number of $\ell$-bins in that dimension. We then define our binned bispectrum estimator in a 3D $\ell$-bin labelled by indices $(a,b,c)$ as: 
\begin{equation}\label{binning}
\BTTg_{abc} = \frac{\sum_{\substack{\ell_{1}\in\Delta_{a}\\\ell_{2}\in\Delta_{b}\\\ell_{3}\in\Delta_{c}}} \left(N_{\Delta}^{\ell_1\ell_2\ell_3} B_{\ell_1\ell_2\ell_3}^{\mathrm{kSZ},\mathrm{kSZ},\delta_g}\right)}{N_{abc}},
\end{equation}
where the estimator normalization $N_{abc} = \sum_{\ell_{1}\in\Delta_{a};\ell_{2}\in\Delta_{b};\ell_{3}\in\Delta_{c}} N_{\Delta}^{\ell_1\ell_2\ell_3}$. Thus, the binned (flat-sky) bispectrum may be considered as a weighted average of all valid $\BTTg_{\ell_1\ell_2\ell_3}$ inside the $\ell$-bin, with the corresponding number of modes $N_{\Delta}^{\ell_1\ell_2\ell_3}$ acting as weights. For the sake of clarity, in Appendix \ref{binflat}, we show that our analytical definition above is consistent with the framework of previous works employing the binned bispectrum approach \cite{bucher_2010, Bucher_2016, Coulton, CoultonSpergel2019}.

\subsubsection{CMB Lensing Contribution}\label{lens_main}
As seen previously \cite{F16} for the existing $\CTTg$ estimator (Section \ref{Clttg}), the binned $\BTTg$ estimator can also potentially receive a contribution due to weak lensing of the CMB (see \cite{Lewis2006}, for example), when it is applied to observed CMB maps. Denoting the unlensed and lensed CMB anisotropies by $\tilde{\Theta} = \Delta\tilde{T}/T_{\mathrm{CMB}}$ and $\Theta = \Delta T/T_{\mathrm{CMB}}$ respectively, we expand up to first order in the lensing potential $\psi$ \cite{Lewis2006}:
\begin{equation}
    \Theta(\mathbf{x}) = \tilde{\Theta}(\mathbf{x}) + \nabla\psi\cdot\nabla\tilde{\Theta}(\mathbf{x}) + ...
\end{equation}
We then perform a calculation analogous to the one in \cite{F16} to derive the lensing correction to our estimator (see Appendix \ref{lens} for details):
\begin{equation}\label{lenscorr}
\Delta\BTTg_{\ell_1\ell_2\ell_3} = -b(\ell_1)b(\ell_2)C_{\ell_3}^{\psi\delta_g}\left(\mathbf{\ell_3}\cdot\mathbf{\ell_1}C_{\ell_1}^{\tilde{T}\tilde{T}}+\mathbf{\ell_3}\cdot\mathbf{\ell_2}C_{\ell_2}^{\tilde{T}\tilde{T}}\right)
\end{equation}
where $C_{\ell}^{\tilde{T}\tilde{T}}$ is the unlensed primary CMB power spectrum. $C_{\ell}^{\psi\delta_g}$ is the angular cross-power spectrum between the lensing potential and the projected-field of galaxies, which can be computed in terms of the lensing convergence $\kappa$ as $C_{\ell}^{\psi\delta_{g}} = (2/\ell(\ell+1))C_{\ell}^{\kappa\delta_{g}}$ \cite{Lewis2006, sherwin}, where
\begin{equation}
    C_{\ell}^{\kappa g} 
    = \int \frac{\mathrm{d}z}{\eta^2(z)}  W_{\kappa}(z)W_{g}(z) P_{\delta\delta}\left(k =\frac{\ell}{\eta(z)}, z \right) \label{GR_Kg}.  
\end{equation}

From Eq.~\eqref{lenscorr}, note that $\Delta\BTTg_{\ell_1\ell_2\ell_3}$ can be positive or negative, depending on the shape of the triangle in $\ell$-space. As shown in Fig. \ref{fig:lens_magnitude}, the \textit{magnitude} of the lensing correction is significant only for triangles having $1500 \lesssim \ell_1, \ell_2 \lesssim 4000$, since lensing dominates in CMB maps at these angular scales. While $|\Delta\BTTg|$ itself peaks for some \Corr{squeezed triangles with a short $\delta_g$ mode (see Fig.\,\ref{fig:lens_magnitude}), it is \textit{relatively} small when compared to the corresponding $\BTTg$ signal in most of these high-SNR bins, as we discuss in Section \ref{fore}.} Moreover, extending the range of $\delta_g$ up to 4200 in Fig. \ref{fig:ext-sliceSO}, we show that the projected-fields kSZ bispectrum and its lensing contribution dominate for different triangle shapes. 
On the other hand, the existing $\CTTg$ estimator compresses not only the full $\BTTg$ signal considered here, but also its lensing correction across all triangle shapes.

\subsubsection{Analytical Covariance Matrix}\label{variance}
As with several bispectra studies done previously (e.g.\,\cite{Bucher_2016, CoultonSpergel2019, Coulton+nGCovar2020}), 
we proceed to use a simplified analytical expression for the covariance matrix of the estimator for forecasting, since this does not require accurate simulated maps for its estimation.  
From Eq.\eqref{defn_alms}, the covariance for a single triplet is given by:
\begin{align}
& \langle\BTTg_{\ell_1\ell_2\ell_3}\BTTg_{\ell_1'\ell_2'\ell_3'}\rangle \nonumber\\
& = \frac{1}{\sqrt{N_{\Delta}^{\ell_1\ell_2\ell_3}N_{\Delta}^{\ell_1'\ell_2'\ell_3'}}}  \sum_{m}\begin{pmatrix}
    \ell_1 & \ell_2 & \ell_3 \\
    m_1 & m_2 & m_3
  \end{pmatrix}\begin{pmatrix}
    \ell_1' & \ell_2' & \ell_3' \\
    m_1' & m_2' & m_3'
  \end{pmatrix} \nonumber\\
& \quad\quad\langle a^{\Theta_b}_{\ell_1 m_1}a^{\Theta_b}_{\ell_2 m_2}a^{\delta_g}_{\ell_3 m_3}a^{\Theta_b *}_{\ell_1' m_1'}a^{\Theta_b *}_{\ell_2' m_2'}a^{\delta_g *}_{\ell_3' m_3'} \rangle.
\end{align}
In the limit of weak non-Gaussianity and using Wick's theorem, the 6-point function of spherical harmonic coefficients above has a decomposition that is dominated by products of three power spectra. From Eq.\eqref{binning}, it follows that in the weak non-Gaussianity limit, the covariance matrix is diagonal, so that the variance of the binned $\BTTg$ bispectrum is:
\begin{align}\label{gauvar}
 V_{abc}^{\mathrm{kSZ},\mathrm{kSZ},\delta_g} = & \frac{\sum_{\substack{\ell_{1}\in\Delta_{a}\\\ell_{2}\in\Delta_{b}\\\ell_{3}\in\Delta_{c}}} \left(N_{\Delta}^{\ell_1\ell_2\ell_3}g_{\ell_1\ell_2\ell_3}C_{\ell_1}^{\Bar{T}\Bar{T},b} C_{\ell_2}^{\Bar{T}\Bar{T},b} C_{\ell_3}^{\delta_g \delta_g} \right)}{N_{abc}^{2}\,f_{\mathrm{sky}}},  
\end{align}
where we have accounted for the effect of the overlapping sky fraction $f_{\mathrm{sky}}$ of the considered surveys. The power spectrum of the observed, cleaned CMB map is computed as (see Section \ref{survey} for details) $C_{\ell}^{\Bar{T}\Bar{T},b} \equiv b^2(\ell) C_{\ell}^{\mathrm{tot}} = b^2(\ell)\left(C_{\ell}^{TT} + C_{\ell}^{\mathrm{kSZ}} + N_\ell\right)$, and the projected galaxy density power spectrum is given by \cite{Limber, extlimber}: 
\begin{equation}
   C_{\ell}^{\delta_g \delta_g} = \int_{0}^{\eta_{\max}} \frac{d\eta}{\eta^2} \, [W^{g}(\eta)]^{2} P_{\delta\delta}\left(k = \frac{\ell}{\eta}, \eta \right) + \frac{1}{\Bar{n}},
\end{equation}
where $\Bar{n}$ is the projected number of galaxies per steradian, appearing in the shot noise term $(1/\Bar{n})$. The symmetry factor $g_{\ell_1\ell_2\ell_3}$ appearing in Eq.\eqref{gauvar} is 2 if the two kSZ multipoles are equal in magnitude (i.e. if $\ell_1 = \ell_2$), and 1 otherwise. 

We now consider the effect of CMB lensing on the estimated covariance matrix of the $\BTTg$ bispectrum. One part of this effect is already accounted for above, given that $C_{\ell}^{TT}$ is the lensed primary CMB power spectrum. Additionally, as pointed out recently \cite{Coulton+nGCovar2020} (in the context of primordial non-Gaussianity), the covariance could potentially receive a non-Gaussian lensing contribution of the form $\langle a^{\Theta_b}_{\ell_1 m_1}a^{\Theta_b}_{\ell_2 m_2}a^{\Theta_b *}_{\ell_1' m_1'}a^{\Theta_b *}_{\ell_2' m_2'}\rangle \langle a^{\delta_g}_{\ell_3 m_3}a^{\delta_g *}_{\ell_3' m_3'} \rangle$. This contribution is sourced by the \textit{connected} trispectrum of CMB temperature, which becomes non-zero in the presence of lensing.

We do a calculation analogous to \cite{Coulton+nGCovar2020} to derive this non-Gaussian contribution to the variance of $\BTTg$ in Appendix \ref{lens} (see Eq.\eqref{lenscorr-var}).
However, we expect this contribution to be relatively small for our analysis with the $\BTTg$ bispectum, because (i) Its overall amplitude due to the multiplicity of possible permutations of $a^{\Theta}$s is lowered by a factor of 9: the average multiplicity is just 4 in our case (as compared to 36 in the context of primordial non-Gaussianity with the primary CMB  $\langle TTT \rangle$ bispectrum in \cite{Coulton+nGCovar2020}), and (ii) for squeezed triangles in the `default' regime assumed in this work (see Sec.\,\ref{analysis}), we find that the lensing contribution to the $\BTTg$ \textit{signal} itself is relatively small (see Fig.\,\ref{fig:sliceSO4}); so, we intuitively expect its variance to receive a similarly small non-Gaussian contribution due to lensing (while also noting the similarity between the expressions in Eqs\,\eqref{lenscorr-var} and \eqref{lenscorr}). Although we have provided its theoretical expression in Appendix \ref{lens} for completeness (Eq.\eqref{lenscorr-var}), we drop this term from our forecasts since it is expected to be small in the regime considered and computationally expensive to evaluate. \Rev{Future work with the $\BTTg$ estimator could potentially include a more accurate estimate of its covariance matrix, including the non-Gaussian contribution due to lensing, by applying our binned bispectrum approach to simulated maps (see e.g. \cite{CoultonSpergel2019}) to quantitatively validate the impact of lensing described in this section.}

\section{Methodology}\label{method}
\subsection{Survey Specifications}\label{survey}
We now summarize the relevant specifications of the CMB and galaxy surveys that we have forecasted for with our novel estimator. Because it utilizes projected-fields of LSS tracers, it requires a CMB map cleaned of contaminating foregrounds, which is obtained by applying some multi-frequency component-separation technique to the observed maps. One such standard technique is the “Internal
Linear Combination” (ILC) method \cite{ILC1, ILC2} which separates the `primary CMB+kSZ' component from foregrounds by minimizing the total variance. 

In this work, we make forecasts for detecting the kSZ by applying this estimator to upcoming CMB data from the \textit{Simons Observatory} (SO) \cite{SO}, and a subsequent potential CMB-S4 experiment \cite{S4}. Both these experiments will map the CMB at a high resolution of $\theta_{\mathrm{FWHM}}\approx1.4$ arcmin (corresponding to an $\ell_{\mathrm{\max}} \approx 8000$), making them well-suited for kSZ studies. To make realistic forecasts, we follow \cite{P23, Bolliet2022} and use post-ILC noise curves for the cleaned CMB maps from SO and CMB-S4, which have been estimated based on simulations and are publicly available\footnote{\label{ilcSO}https://github.com/simonsobs/so\_noise\_models/tree/master/\newline
LAT\_comp\_sep noise/v3.1.0; standard ILC:deproj-0}${}^{,}$\footnote{\label{ilcS4}https://sns.ias.edu/jch/S4\_190604d\_2LAT\_Tpol\_default\newline\_noisecurves.tgz}. These noise power spectra $N_{\ell}$ are considerably larger than the detector-only white noise (e.g. see Fig. 4 in \cite{RS_SF}) since they also include any residual foregrounds still present after ILC; forecasted SNRs for the related $\CTTg$ estimator were found to be a factor of $\sim3$ smaller with these realistic post-ILC noise curves, as compared to the detector-only case \cite{P23}.  

\begingroup
\begin{table}[t]
\begin{center}
\renewcommand{\arraystretch}{1.28} 
\begin{tabular}{|c|c|c|c|c|c|}
\hline
\hline
CMB&$\theta_{\mathrm{FWHM}}$&Noise&$\ell_{\mathrm{\min}}$&$\ell_{\mathrm{\max}}$&$f_{\mathrm{sky}}$ \\
 experiment&[arcmin]&model&& & \\ \hline
SO&1.4&post-ILC$^{\ref{ilcSO}}$&100&8000&0.4\\
CMB-S4&1.4&post-ILC$^{\ref{ilcS4}}$&100&8000&0.4\\
\hline
\hline
\end{tabular}
\end{center}
\caption{Specifications for the CMB experiments forecasted for in this work. We use realistic post-ILC noise curves for \textit{Simons Observatory} (SO) \cite{SO} and CMB-S4 \cite{S4}, which account for the beam implicitly and also include residual foreground contamination in the cleaned maps. Note that the estimated noise levels of CMB-S4 are considerably lower than that of SO (see Fig. 1 in \cite{Bolliet2022}, for example).}
\label{tab:cmb_specs}
\end{table}
\endgroup

Throughout this work, the unlensed ($C_{\ell}^{\tilde{T}\tilde{T}}$) and lensed ($C_{\ell}^{TT}$) primary CMB power spectra are computed with CAMB \cite{camb1, camb2} for the fiducial $\Lambda$CDM model defined by the base parameters:
$\{H_{0}, \Omega_{b}h^{2}, \Omega_{c}h^{2}, 10^{9}A_{s}, n_{s}, \tau_{re} \}$, 
using their best-fit \textit{Planck}-2018 values (rightmost column of Table 1 of \cite{Planck2018}). The total power spectrum of the post-ILC CMB map is $C_{\ell}^{\mathrm{tot}}$ = $\left(C_{\ell}^{TT} + C_{\ell}^{\mathrm{kSZ}} + N_\ell\right)$; it is computed to estimate the observable's noise covariance (Eq.\,\eqref{gauvar}) by using a template (theoretical) kSZ power spectrum\footnote{https://github.com/nbatta/SILC/blob/master/data/ \\
ksz\_template\_battaglia.csv} derived from hydrodynamical simulations \cite{Battaglia2010}.

As mentioned in Section \ref{theory}, we consider a projected-field of galaxies as our LSS tracer here. We make forecasts for SO and CMB-S4 by combining with the existing WISE \cite{WISE} photometric galaxy catalog consisting of more than 500 million objects. Following previous works \cite{F16, Hill2016, P23}, we similarly select a sample of 46.2 million galaxies, based on the same criterion as \cite{WISE2, WISE3}. The overlapping sky fraction between WISE and these CMB surveys is $f_{\mathrm{sky}} \approx 0.4$. The redshift distribution of this sample peaks at $z \approx 0.3$ and extends up to $z = 1$ \cite{WISE_z} (see Fig. 10 of \cite{F16}). 

\subsection{Analysis Choices}\label{analysis} 
We make a few key analysis choices while implementing the projected-fields kSZ bispectrum as introduced in Section \ref{ourest}. To begin with, we use broad redshift bins of width $\Delta z = 0.1$ since the estimator only requires a rough $z$-distribution of the galaxies; all LOS integrals (including Eq. \eqref{bis}) are thus replaced by sums spanning the redshift range ($z<1$ here). Following earlier works \cite{F16, P23}, we only consider modes with $\ell>\ell_{\mathrm{\min}}\equiv100$ in all 3 fields, to avoid picking up an additional signal from the (linear) ISW effect that is expected to be significant at very large ($\ell<100$) scales \cite{WISE2, ISWdet_2016, F16}. 

Unlike the existing implementation employing the $\CTTg$ estimator \cite{Hill2016, F16, Kusiak2021, Bolliet2022, P23}, we \textit{choose}  to use an $\ell$-cut in the LSS field, without restricting the $\ell$-range of kSZ modes (which dominate at smaller scales $\ell\gtrsim4000$). In our `default' analysis, we choose a conservative $\ell_{\mathrm{\max}}$ of 700 for the $\delta_g$ field, restricting to large scales where the galaxy density field is expected to be well-modeled by a linear galaxy bias $b_g$ (see e.g.\cite{unwise, Farren+2024}). The best-fit value for WISE galaxies is $b_{g} = 1.13$, as determined earlier using \textit{Planck} maps \cite{F16, Hill2016}. Working within the linear regime for $\delta_g$ enables us to circumvent complications due to the HOD and non-linear galaxy bias terms, which makes our results more robust.

For binning the predicted $\BTTg$ signal (Section \ref{ourest}), we choose a uniform bin width of $\Delta\ell = 100$ in each of the three harmonic dimensions, which is broad enough so that there are a sufficient number of bins that contain a non-zero number of triangle modes within them. 

\subsection{Numerical Implementation}\label{num} 
Accurately predicting the $\BTTg$ signal involves computing the underlying key statistic, $\Blos$ (Eq.\,\eqref{bis}) using its improved model \cite{P23} given by Eqs.\,(A1-A5). Following \cite{D05,P23} and several earlier works on the kSZ, we make the simulation-based phenomenological substitution \cite{Zhang2004} for each velocity field appearing in these equations; the non-linear matter density field $\delta_m$ is substituted into the linear continuity equation: 
\begin{equation}\label{lin-vel}
   \mathbf{v}(\mathbf{k}) = i\frac{faH\delta_m(\mathbf{k})}{k} \hatbf{k},
\end{equation}
which is a good description of peculiar velocities, since most of their power comes from large scales in the linear regime. Here, $a, f$, and $H$ are the scale factor, linear growth rate, and Hubble parameter at that redshift respectively. The resulting non-linear matter power spectra $P_{\delta\delta}$ are calculated for the fiducial $\Lambda$CDM model with CAMB and the standard Halofit prescription \cite{Takahashi2012}. 

In Section \ref{fish-mnu}, we will consider an extension to the $\Lambda$CDM model with another free parameter: the sum of neutrino masses ($\Sigma m_{\nu}$), which is known to be non-zero from flavor oscillation detections \cite{super-Kamiokande, SNO}. These independent experiments permit for both, a normal hierarchy of neutrinos ($m_1$\,$\simeq$\,$m_2$\,$\simeq$\,0\,$\ll$\,$m_3$; with a minimum $\mnus$ of $\sim58$ meV), and an inverted hierarchy ($m_3\ll m_1$\,$\simeq$\,$m_2$; with a minimum $\mnus$ of $>100$ meV) \cite{mnu_CMBLSS, mnu_whitep}. In our forecasts, we assume a fiducial value of 60 meV. The presence of massive neutrinos suppresses the clustering of matter at scales smaller than their free-streaming scale, which is defined as $k_{\mathrm{fs}}(z) = 0.018 \,\Omega_{m}^{1/2}(z)[m_{\nu}$/1 eV] h Mpc$^{-1}$.(e.g. \cite{Lesgourgues, mnu_whitep}). They also introduce a small scale-dependence in the growth rate $f$, which has two powers appearing in our estimator (from Eq.\eqref{lin-vel}). This dependence is well-modeled by a fitting function \cite{Kiakotou2007}: $f(k, z) \approx \mu(k)\Omega_{m}^{\alpha}(z)$ with $\alpha\approx0.55$ for $\Lambda$CDM, and
\begin{equation}\label{mnu-f}
    \mu(k) = 1 - A(k)\Omega_{\Lambda}f_{\nu} + B(k)f_{\nu}^{2} - C(k)f_{\nu}^{3},
\end{equation}
where the coefficients $A(k), B(k), C(k)$ are obtained via interpolation (see Table II in \cite{Kiakotou2007}). $f_{\nu} =\Omega_{\nu}/\Omega_{m}$ is the fractional contribution of neutrinos to the total matter density ($\approx$ 0.0045 fiducial value), and $\Omega_{\Lambda}$ is the dimensionless dark energy density. 

The non-linear matter bispectrum $B_{\delta\delta\delta}$ is computed by following earlier works \cite{F16, Hill2016, Kusiak2021, P23} and using the following fitting function \cite{Gilmarin2012} for simplicity:
\begin{equation}\label{gil}
    B_{\delta\delta\delta}(k_1, k_2, k_3) = \sum_{cyc} 2F_{2}(k_1, k_2, k_3)P_{\delta\delta}(k_1)P_{\delta\delta}(k_2),
\end{equation}
where $\sum_{cyc}$ denotes a cyclic sum over $(k_1, k_2, k_3)$. The scale and redshift dependence of the kernel $F_2$ are fitted using $\Lambda$CDM-only N-body simulations, although the formula itself is general enough to be applicable at the first-order for certain $\Lambda$CDM-extensions, due to the weak cosmological dependence of $F_2$ \cite{Gilmarin2012}. The effect of massive neutrinos on the matter bispectrum is via the suppression of the matter power spectrum, which is consistent at first order with results from second-order perturbation theory \cite{vlah} and simulations \cite{Ruggeri2017, Coulton}.

While the binned bispectrum approach is relatively fast when applied directly to maps, in the absence of accurate simulated maps, we calculate the theoretical prediction by summing over all possible triplets (Eq.\eqref{binning}). In practice, since the projected-fields kSZ bispectrum is smooth, we speed up the process within each $\ell$-bin indexed by $(a,b,c)$ by directly computing $\BTTg$ for multipole-triplets separated by a (3D) sub-binwidth of $\sim10$, and summing over the weights ($N_{\Delta}^{\ell_1\ell_2\ell_3}$) within each sub-bin to obtain the final binned signal $\BTTg_{abc}$. The numerical computation of the statistic $\Blos$ in each redshift bin is sped up by employing the same emulators\footnote{\url{https://github.com/dylancromer/ostrich}} \cite{ostrich_cite} based on Gaussian Process Interpolation that were used in the previous work on $\CTTg$ \cite{P23}. 

\section{Forecasts}\label{fore} 
\begin{figure*}
\center
{
\includegraphics[width=0.84\textwidth]{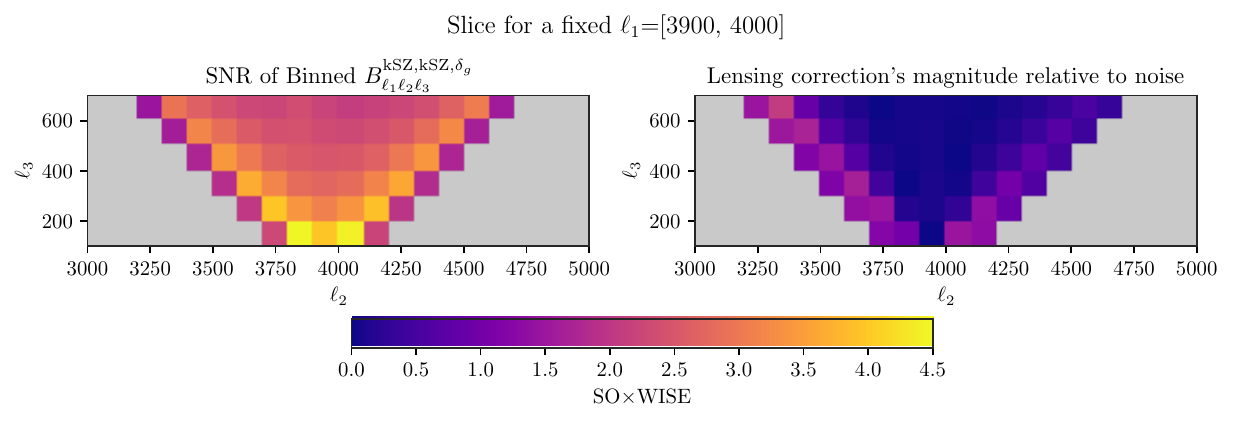} 
\includegraphics[width=0.84\textwidth]{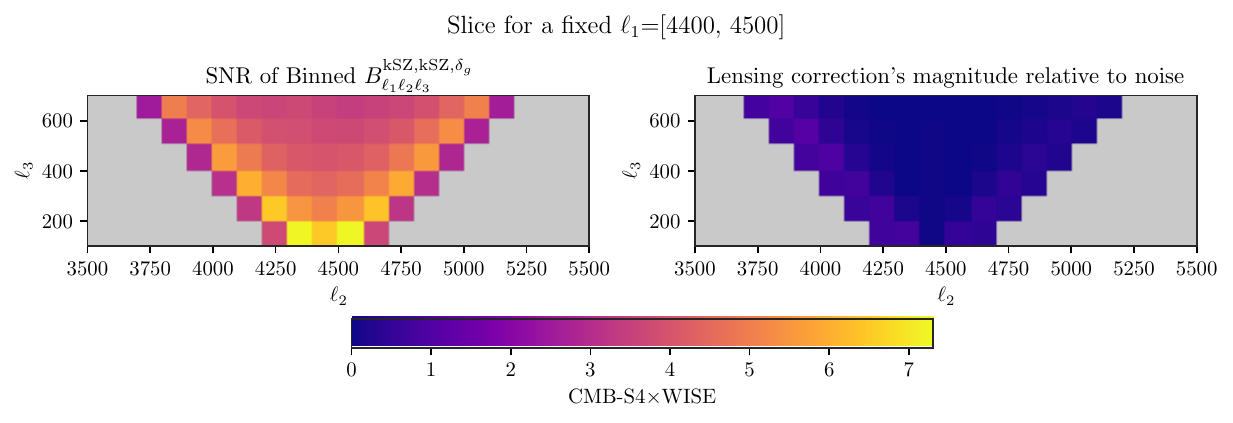} 
}
\caption{Examples of 2D slices in 3D harmonic space in the default scenario with a fixed $\ell_1$-bin of [3900,4000] and [4400,4500] for SO$\times$WISE (top-left) and CMB-S4$\times$WISE (bottom-left) respectively, showing the individual signal-to-noise-ratios (SNR) of $\BTTg$ within each 3D bin. \Corr{The absolute value of the lensing contribution relative to the noise covariance in each bin $\left(|\Delta\BTTg|/\sqrt{V^{\mathrm{kSZ},\mathrm{kSZ},\delta_g}}\right)$ is displayed for SO$\times$WISE (top-right) and CMB-S4$\times$WISE (bottom-right) respectively using the same colorbars as their corresponding SNR plots.} The grey bins do not contain any $\ell$-triplet satisfying the triangle inequality (i.e. no signal under isotropy).}  
\label{fig:sliceSO4}
\end{figure*}
\begingroup
\begin{table*}
\centering
\setlength{\tabcolsep}{3.4pt}
\renewcommand{\arraystretch}{1.28} 
         \begin{tabular}{|c|c|c|c|c|c|c|}
         \hline
         \hline
 &\multicolumn{2}{|c|}{$B^{\mathrm{kSZ},\mathrm{kSZ},\delta_g}$ (default)}  &\multicolumn{2}{|c|}{$B^{\mathrm{kSZ},\mathrm{kSZ},\delta_g}$ (extended)}
 &\multicolumn{2}{|c|}{$C_{\ell}^{\mathrm{kSZ}^{2}\times\delta_{g}}$ (from \cite{P23})}\\   \hline
 \multicolumn{1}{|c|}{$\ell_{\mathrm{max}}$ of $\delta_g$ used=} &\multicolumn{2}{|c|}{700}&\multicolumn{2}{|c|}{4200}&\multicolumn{2}{|c|}{8000}\\   \hline
 &SO $\times$ WISE&CMB-S4$\times$ WISE&SO $\times$ WISE&CMB-S4$\times$ WISE&SO $\times$ WISE&CMB-S4$\times$ WISE \\   \cline{2-7}
 \multicolumn{1}{|c|}{$\mathrm{SNR_{tot}}$}& 106 & 200 & 221 & 418 & 113 & 127 \\ \hline\hline 
 \end{tabular} 
 \caption{Forecasted signal-to-noise ratios (SNR) of SO$\times$WISE and CMB-S4$\times$WISE for detections with the projected-fields kSZ bispectrum (this work), where the LSS field is restricted to linear scales ($\ell_{\mathrm{max}}=700$) in the default scenario, and up to $\ell_{\mathrm{max}}=4200$ in the extended scenario. We compare these against the corresponding SNRs with the related kSZ$^{2}\times\delta_g$ estimator, as forecasted in \cite{P23} with an effective $\ell_{\mathrm{max}}=8000$.}
 \label{table:snrs}
 \end{table*}
 \endgroup
\begin{figure*}
\center
{
\includegraphics[width=1.04\textwidth]{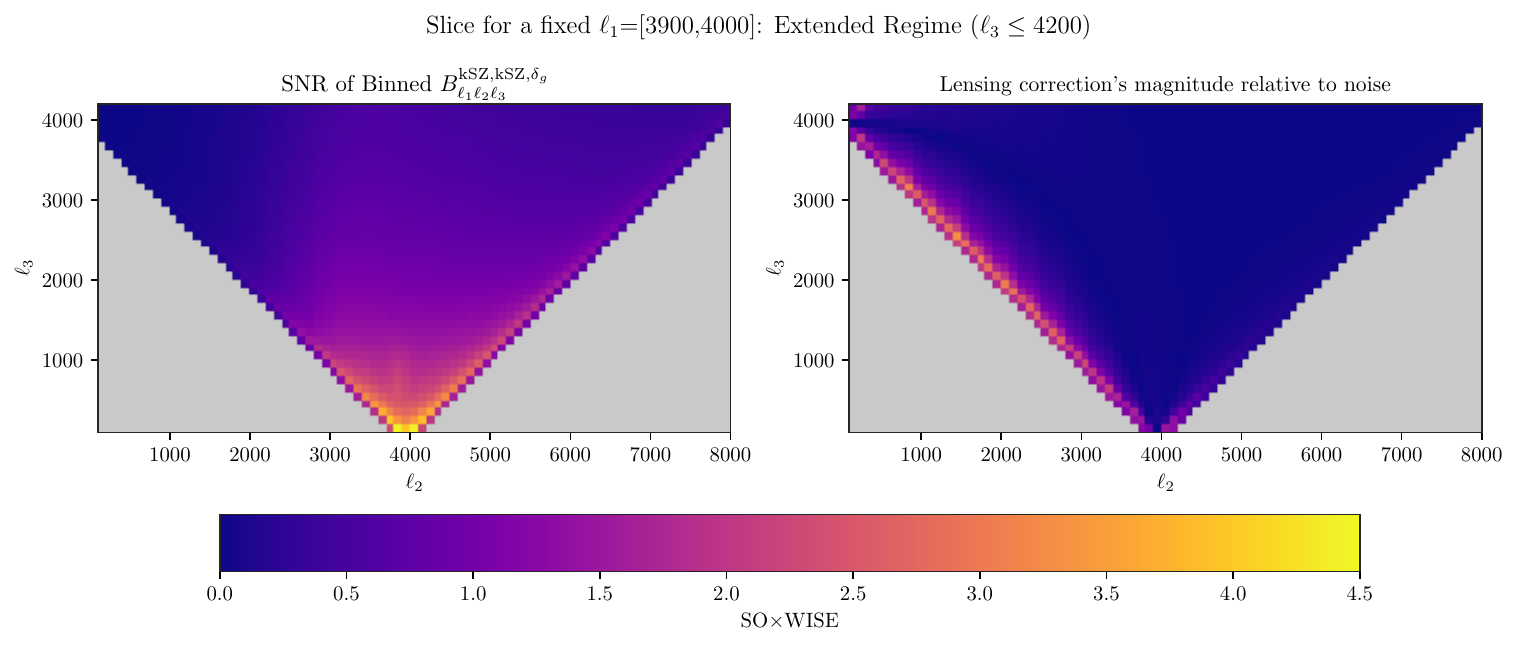} 
}
\caption{A 2D slice in 3D harmonic space in the extended scenario with a fixed $\ell_1$-bin of [3900,4000] showing the individual SNR ($= \left(\BTTg/\sqrt{V^{\mathrm{kSZ},\mathrm{kSZ},\delta_g}}\right)$ within each bin (left), and \Corr{the absolute value of the lensing contribution relative to the noise covariance in each bin $\left(|\Delta\BTTg|/\sqrt{V^{\mathrm{kSZ},\mathrm{kSZ},\delta_g}}\right)$ (right)} for SO$\times$WISE using the same colorbar. The SNR peaks for squeezed triangles (of the type in Fig.\,1), with a maximum of $\sim4.5$ for a bin within the slice shown here, \Corr{while the \textit{relative} magnitude of the lensing correction peaks for different triangle shapes.}} 
\label{fig:ext-sliceSO}
\end{figure*}

We now present forecasts for detecting the kSZ effect using the
novel projected-fields binned bispectrum (as defined in Section \ref{ourest}) with upcoming CMB data from SO or CMB-S4 and existing photometric galaxy data from WISE. We proceed to numerically compute the predicted signal $\BTTg_{abc}$ in bins with 3D binwidth $\Delta\ell=100$, using the survey specifications and methodology detailed earlier. Under weak non-Gaussianity, the covariance matrix of this estimator is diagonal (in 3D), and the variance $V_{abc}^{\mathrm{kSZ},\mathrm{kSZ},\delta_g}$ is thus given by Eq.\eqref{gauvar}. 

As mentioned in Section \ref{analysis}, in our analysis, we use an $\ell_{\mathrm{\max}} = 700$ by default for the $\delta_g$ modes, in order to stay within the linear galaxy bias regime. Due to this restriction, only around $(1/8)^{\mathrm{th}}$ of the total number of bins contain a non-zero number of closed triangles (such that $(\ell_1,\ell_2,\ell_3)$ satisfy the triangle inequality), and thus have a non-zero signal. For each such bin indexed by $(a,b,c)$, we compute its individual signal-to-noise ratio, SNR$_{abc} = \left(\BTTg_{abc}/\sqrt{V_{abc}^{\mathrm{kSZ},\mathrm{kSZ},\delta_g}}\right)$. 

Since the harmonic space is 3D, for easy visualization, we show the SNRs for bins in 2D `slices' that span the entire $\ell_2$ and $\ell_3$ range and have a fixed $\ell_1$-bin. We find that the individual SNR for SO$\times$WISE peaks at a value of $\sim4.5$ for a bin in the $\ell_1=[3900,4000]$ slice, while that for CMB-S4$\times$WISE peaks at a value of $\sim7$ for a bin in the $\ell_1=[4400,4500]$ slice. This is expected because the kSZ effect begins to dominate in CMB maps at small scales ($\ell>3500$), and because the post-ILC noise levels in CMB-S4 will be lower than that of SO (see Fig. 1 in \cite{Bolliet2022}, for reference). Both these slices are shown in the left panels of Fig. \ref{fig:sliceSO4}. We find that across all 2D slices, the SNR peaks for the most `squeezed' triangle configurations which have two long sides ($\ell_1$ and $\ell_2$, since the kSZ is more dominant at smaller scales) and one very short side ($\ell_3$, the $\delta_g$ mode), as suggested earlier with the kSZ$^2$ estimator \cite{Kusiak2021, Bolliet2022}. For such squeezed triangle types (shown schematically in Fig.\,1), it is even more crucial to use the improved model of $\Blos$ \cite{P23} in our forecasts (instead of the approximate one \cite{Dore2004, D05}), because the `extra-3' term (given by Eq.\eqref{extra-3}) contributes significantly to the predicted signal.

We also forecast cumulative SNRs by combining across bins spanning the entire 3D $\ell$-range, given by (e.g. \cite{Tegmark1997}): 
\begin{equation}\label{SNR}
\mathrm{SNR_{tot}} = \sqrt{\sum_{a,b,c}\frac{\left(\BTTg_{abc}\right)^2}{V_{abc}^{\mathrm{kSZ},\mathrm{kSZ},\delta_g}}}. 
\end{equation}
We find a total SNR of $\sim106$ and $\sim200$ for SO$\times$WISE and CMB-S4$\times$WISE, in the default scenario where we only keep the linear $\delta_g$ modes with $\ell<700$.

Now, the existing projected-fields $\CTTg$ estimator compresses all possible triangle shapes, so that effectively, the $\delta_g$ modes it includes have an $\ell_{\mathrm{max}}$ that is the same as the estimator's maximum multipole, which has been $\sim8000$ in previous forecasts for SO and CMB-S4 \cite{F16,Bolliet2022,P23}. To compare the information content of this `kSZ$^{2}\times\delta_g$' estimator and the $\BTTg$ estimator, we temporarily consider an `extended' case where we apply our estimator with $\delta_g$ modes extended up to an $\ell \sim4200$. The corresponding $\ell_1=[3900,4000]$ slice for SO$\times$WISE is visualized in the left panel of Fig.\,\ref{fig:ext-sliceSO}. Even with this extended range, the $\BTTg$ bispectrum has its highest detection significance with squeezed triangle shapes that have a short galaxy mode (Fig.\,1). These high-SNR triangles are retained even if we restrict our estimator to the linear galaxy bias regime (default case), and unlike the $\CTTg$ estimator, their information is not mixed with other triangle configurations.   
 
Table \ref{table:snrs} shows that the forecasted total SNR of our estimator nearly doubles for both SO$\times$WISE and CMB-S4$\times$WISE when we extend the range of selected $\delta_g$ modes from an $\ell_{\mathrm{max}}$ of 700 to 4200, indicating that a non-negligible amount of signal is present in these lower-SNR triangles as well. Since it is a full bispectrum, this estimator contains more information than the compressed kSZ$^{2}\times\delta_g$ estimator, so that its cumulative SNR in the extended case (with $\ell_{\mathrm{max}} = 4200$) is higher than that of $\CTTg$ (effective $\ell_{\mathrm{max}} = 8000$). 

In the right panel of Fig.\,\ref{fig:ext-sliceSO}, we visualize the absolute value of the lensing contribution divided by the noise covariance for each bin in that slice - i.e. $\left(|\Delta\BTTg|/\sqrt{V^{\mathrm{kSZ},\mathrm{kSZ},\delta_g}}\right)$ - for SO$\times$WISE, using the same colorbar as the SNR plot (left panel). In this extended regime ($\ell_3 \lesssim 4200$), the magnitude of the \Corr{lensing correction is significant compared to the $\BTTg$ signal only for triangles that have both CMB modes $\ell_1, \ell_2 \lesssim$ 4000; these are the angular scales at which lensing is important in CMB maps.} Thus, we find that the SNR of $\BTTg$ and the relative magnitude of $\Delta\BTTg$ peak for different shapes of triangle configurations. Specifically, \Corr{in the default scenario assumed throughout this work ($\ell_3 \lesssim 700$), the lensing correction is non-negligible but small \textit{relative} to the $\BTTg$ signal for most squeezed triangles, as shown for zoomed-in slices in the right panels of Fig.\,\ref{fig:sliceSO4}.}

We continue to choose the default scenario of a conservative $\ell_{\mathrm{max}} = 700$ for the $\delta_g$ mode throughout the rest of this work. 
Furthermore, as discussed above, even after galaxy modes are restricted to the linear regime, the highest-SNR triangles with squeezed shapes are retained, leading to a high cumulative detection significance.

\section{Cosmological Dependence}\label{cosmo-sec}
\begin{figure*}
\center
{
\includegraphics[width=0.96\textwidth]{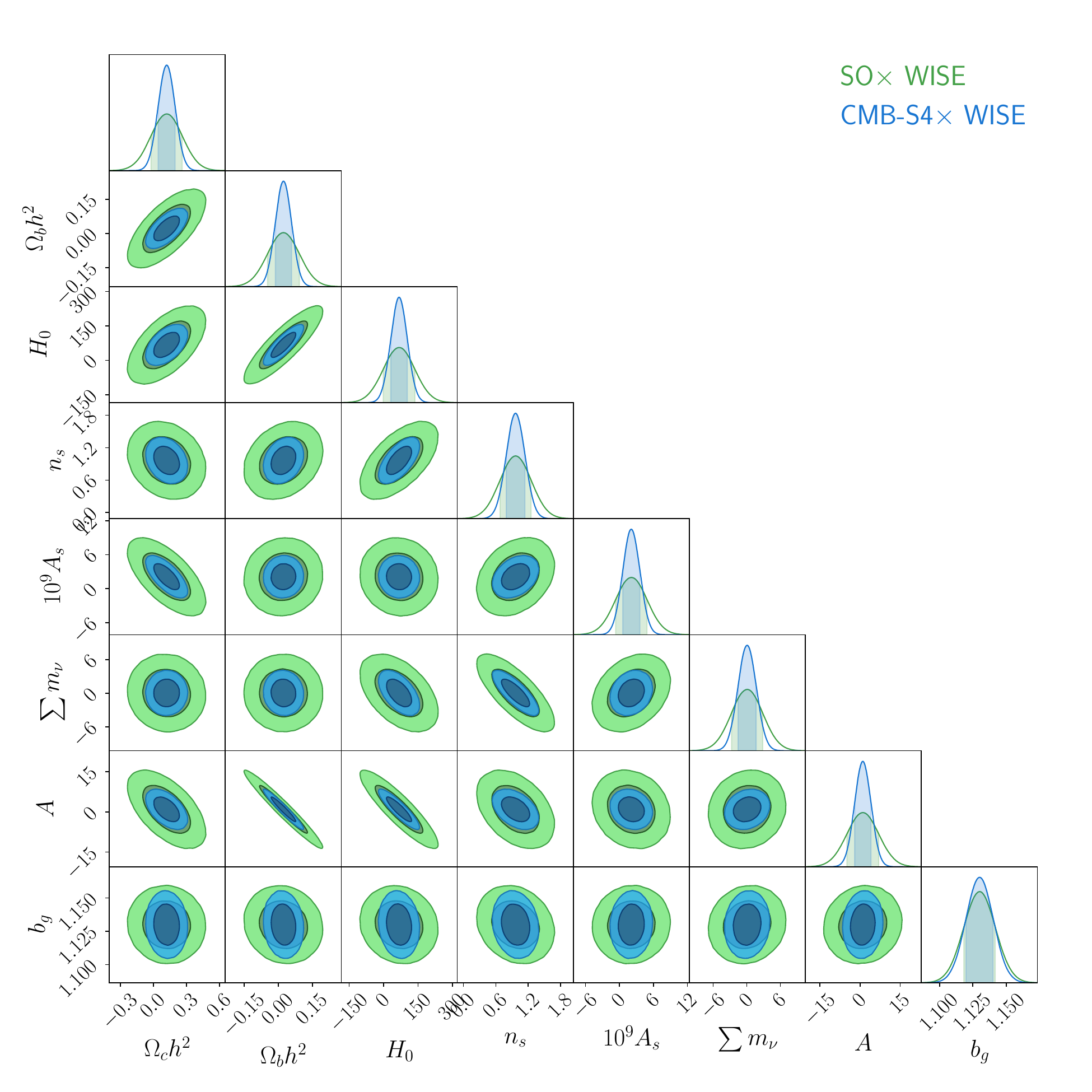} 
}
\caption{Fisher forecasts for marginalized errors and \Corr{degeneracies among $\Lambda$CDM parameters, $\mnus$ (in eV here), $A$, and $b_g$ using the projected-fields kSZ bispectrum alone}, without assuming \textit{any} external priors, and with realistic post-ILC noise for SO and CMB-S4. Contours show the 2D marginalized 68$\%$ (darker) and 95$\%$ confidence ellipses for SO$\times$WISE (green) and CMB-S4$\times$WISE (blue).}
\label{fig:Fisher-noA}
\end{figure*}

We forecasted that the projected-fields kSZ bispectrum will have high detection significance (SNR = 100-200) when applied to upcoming CMB data from SO or CMB-S4, in combination with photometric galaxy data from a survey such as WISE. While restricting to the linear galaxy bias regime, this kSZ signal could potentially be used as a probe of the baryon abundance (i.e.\,$f_b$) or the baryon density profile in the late-time Universe, while being immune to modeling uncertainties in the HOD or non-linear galaxy bias at smaller scales. The astrophysical interpretation of the signal will also be limited by residual uncertainty in the fiducial $\Lambda$CDM parameters that govern it - the familiar `optical-depth degeneracy' present for all kSZ estimators (e.g.\cite{Bhattacharya2008, Smith2018, P23}). 

To this end, we now study the cosmological dependence of the novel $\BTTg$ signal. The baseline set of parameters that we allow to vary is given by:
\begin{equation}\label{thetavec}
\pmb{\theta}_{\mathrm{min}} = \{H_{0}, \Omega_{b}h^{2}, \Omega_{c}h^{2}, 10^{9}A_{s}, n_{s}\}+\{\Corr{A, b_g}\},
\end{equation}
where $\BTTg$ and $\Delta\BTTg$ both scale linearly with $b_g$, the linear galaxy bias (from Eqs.\eqref{bis},\eqref{lenscorr}). $A$ is an astrophysical nuisance parameter that acts as a free, \emph{scale-independent} amplitude of $\BTTg$, with a fiducial value of 1; it is defined similar to previous works that characterized and constrained the free amplitude of the compressed $\CTTg$ signal \cite{Hill2016, F16, Kusiak2021, Bolliet2022, P23}. The other five are baseline $\Lambda$CDM parameters with fiducial values assumed to be the best-fit values from \textit{Planck}-2018 \cite{Planck2018}. We do not explicitly model the dependence of the $\BTTg$ signal on $\tau_{re}$, the optical depth to reionization (which is fixed at its best-fit \textit{Planck}-2018 value). This is because we are only considering a low-$z$ ($< 1$) galaxy sample in our forecasts here, and the late-time kSZ effect thus extracted is not directly sensitive to $\tau_{re}$. We do include $\tau_{re}$ in all external cosmological priors that we use, and marginalize over it throughout our analysis.   

To develop a basic understanding, we first consider only the dependence of the $\BTTg$ signal's \textit{amplitude} on each of the five $\Lambda$CDM parameters. We follow \cite{P23} and roughly model the dependence
on each parameter as a power-law scaling within each 3D $\ell$-bin. As detailed in Appendix \ref{cosmodep}, we find that the amplitude is highly sensitive to $\Omega_{c}h^{2}$ and $A_{s}$ in particular, which largely determine the derived parameter $\sigma_{8}$'s value. Similar to the compressed $\CTTg$ signal \cite{Dore2004, P23}, from Eq.\eqref{gil}, the overall amplitude roughly scales as: $\BTTg\propto P_{\delta\delta}^{3}\propto \sigma_{8}^{6-7}$, thus explaining this sensitivity. This simplistic picture is only to provide some intuition about the signal's overall amplitude, and does not fully reflect the scale-dependence induced by cosmology. 
 
We now use the Fisher matrix formalism (e.g.\,\cite{Tegmark1997, Amendola_DE}) to further study the dependence of the $\BTTg$ signal on these parameters. The Fisher matrix here is given by:
\begin{align}\label{fish}
&F_{ij} = \sum_{\substack{a,a'\\b,b'\\c,c'}}\left(\frac{\partial{\BTTg_{abc}}}{\partial{\theta_{i}}}\right) [M^{-1}]_{abc,a'b'c'} \left(\frac{\partial{\BTTg_{a'b'c'}}}{\partial{\theta_{j}}}\right) \nonumber \\ 
&= \sum_{a,b,c}\frac{1}{V_{abc}^{\mathrm{kSZ},\mathrm{kSZ},\delta_g}}\left(\frac{\partial{\BTTg_{abc}}}{\partial{\theta_{i}}}\right)  \left(\frac{\partial{\BTTg_{abc}}}{\partial{\theta_{j}}}\right),
\end{align}
where we have simplified the formula by substituting the covariance matrix $M$ of the binned bispectrum under weak non-Gaussianity, given by $V_{abc}^{\mathrm{kSZ},\mathrm{kSZ},\delta_g}$ (see Section \ref{ourest}). Partial derivatives of the estimator are estimated numerically using the same step sizes as \cite{cromer_table} (Table I therein) for the $\Lambda$CDM parameters, and are checked for numerical stability. 

To begin with, if the goal is to use the $\BTTg$ estimator to constrain the baryonic abundance while assuming a fixed cosmology (assuming fiducial $\Lambda$CDM values), the parameter set reduces to: $\pmb{\theta} = \{A, b_g\}$. Similar to previous works with the kSZ$^2$ estimator \cite{Hill2016, F16, Kusiak2021}, since $\BTTg\propto f_b^2$, the baryon fraction $f_b$ can be inferred directly by constraining the \Rev{free amplitude $A$. Note that equivalently, $A$ is proportional to the square of the integrated optical depth ($\tau$) of this kSZ measurement}, which receives electron density contributions from redshifts spanned by the galaxy sample used (here, WISE spans $z<1$); \Rev{this $\tau$ is distinct from the total optical depth to reionization parameter ($\tau_{\mathrm{re}}$).} 
\Corr{While the kSZ signal $\BTTg$ is also $\propto b_g$, the presence of the lensing correction $\Delta\BTTg \propto b_g$ breaks some of the degeneracy between $A$ and $b_g$.} As shown in the left panel of Fig.\,\ref{fig:A-bg} in Appendix \ref{cosmodep}, we forecast $\sim 1\%$ $1\sigma$ errors on both $A$ and $b_g$ by jointly constraining them with SO$\times$WISE and CMB-S4$\times$WISE using this estimator alone.

Next, suppose we include the dependence of the $\BTTg$ signal on cosmological parameters as well (Eq.\eqref{thetavec}). We assume an external template for the lensing correction $\Delta\BTTg$ using the fiducial $\Lambda$CDM model, while explicitly modeling and including its linear dependence on $b_g$. Performing a Fisher analysis and marginalizing over $\Lambda$CDM parameters, we find that $b_g$ continues to be well-constrained (at $\sim1\%$) because of the lensing correction's distinct scale-dependence. However, opening up the cosmological parameter space leads to a large degradation of joint constraints on $A$, as shown in Fig.\,\ref{fig:A-bg} (right). This is because \Corr{for triangles within the default regime (i.e., with linear $\delta_g$ modes $< 700$), $A \propto f_b^2$ is highly degenerate with $\Omega_bh^2$ (and $H_0$) in particular among the $\Lambda$CDM parameters. Therefore, for cosmological applications of the $\BTTg$ bispectrum, we first estimate the scale-independent amplitude $A$ for the same datasets by using the projected-fields $\CTTg$ estimator, which compresses the $\BTTg$ signal across all triangle shapes in an extended regime; we thus include a $7\%$ Gaussian prior on $A$ \emph{alone}} (as forecasted with $\CTTg$ after marginalizing over $\Lambda$CDM parameters in \cite{P23}). Note that we do not combine \emph{any} cosmological information from $\CTTg$ in our Fisher analysis; we only use it \Rev{to account for the `optical depth degeneracy' of $\BTTg$ by including its estimate of the mutual scale-independent astrophysical amplitude $A$, which scales $\propto \tau^2$.} 

\subsection{Forecasted Constraints including Neutrino Masses}\label{fish-mnu} 
\begingroup
\begin{table*}
\centering
\setlength{\tabcolsep}{3.5pt}
\renewcommand{\arraystretch}{1.28} 
         \begin{tabular}{|c|c|c|c|c||c|c|}
         \hline
         \hline
 &\multicolumn{4}{|c||}{$\BTTg$ + primary CMB prior ($\Lambda$CDM)} 
 &\multicolumn{2}{|c|}{\multirow{2}{*}{$\BTTg$ + primary \textit{Planck} + DESI BAO}   }\\ \cline{2-5}  
 &\multicolumn{2}{|c|}{\textit{Planck}} 
 &\multicolumn{2}{|c||}{(SO/CMB-S4 + LiteBIRD)} 
 & \multicolumn{2}{|c|}{}
 \\   \hline   
 \multirow{2}{*}{\stackanchor{$ \sigma(\Sigma m_{\nu})$}{[meV]}} &SO $\times$ WISE&CMB-S4$\times$ WISE&SO $\times$ WISE&CMB-S4$\times$ WISE&~~~~~~SO $\times$ WISE~~~~~~&CMB-S4$\times$ WISE\\\cline{2-7} 
 & 159 & 129 & 97 & 82 & 54 & 53\\ \hline\hline 
 \end{tabular} 
 \caption{\Corr{Forecasted 1$\sigma$ errors on $\mnus$ with the projected-fields kSZ bispectrum} for SO$\times$WISE and CMB-S4$\times$WISE in the default regime where galaxy modes are restricted to linear scales. Marginalized constraints on $\mnus$ are obtained using a primary CMB prior on $\Lambda$CDM parameters from \textit{Planck} [left columns] or a (high-$\ell$ SO/CMB-S4 + low-$\ell$ LiteBIRD) prior [middle columns]. \Rev{Assuming a prior on $\Lambda$CDM+$\mnus$ from primary \textit{Planck} data and forecasted BAO observations with DESI, addition of information from the $\BTTg$ signal improves the $\sigma(\mnus)$ constraint from 75 meV to about 54 meV} [right columns]. A Gaussian prior on the free amplitude $A\propto\tau^2$ is applied as explained in the text, to account for the optical-depth degeneracy.  
 }
 \label{table:paramserror-ksz+pl}
 \end{table*}
 \endgroup
We now illustrate the potential of the projected-fields kSZ bispectrum as a probe of cosmology by considering a specific extension of the $\Lambda$CDM model: we allow the sum of neutrino masses $\mnus$ to vary as an additional parameter.  
We again restrict the galaxy modes to large scales only (Section \ref{analysis}), so that the linear galaxy bias $b_g$ is a sufficiently accurate description in this regime. As explained above, we also include $A$, the overall free amplitude of the $\BTTg$ signal, as a scale-independent astrophysical nuisance parameter, \Rev{thus accounting for the `optical depth degeneracy' here. However, note that the forecasted constraints here can only be achieved if other parameters governing the baryon density profile can be assumed to be known precisely enough externally (e.g. via a joint analysis with other observables, such as velocity-weighted stacking kSZ measurements \cite{Schaan2021, Hadzhiyska2024}), since they have a scale-dependent impact on the $\BTTg$ signal at non-linear scales.} 

Neutrino masses impact the $\BTTg$ signal through their suppression of clustering of matter at small scales, as well as their induced scale-dependence in the growth rate of matter ($f$) occurring at intermediate scales; both these effects are numerically implemented as detailed in Section \ref{num}. We initially perform a Fisher analysis without assuming any prior at all. 2D contours in Fig. \ref{fig:Fisher-noA} show the estimated 1$\sigma$ and 2$\sigma$ marginalized uncertainties for each pair of parameters. As expected, the lower noise levels of CMB-S4 give much tighter constraints with CMB-S4$\times$WISE than with SO$\times$WISE. Among the contours displayed, the strongest degeneracy is between $A$ and $\Omega_b h^2$, since both these parameters linearly scale the amplitude of the $\BTTg$ signal for squeezed triangles.

Next, we apply an external prior on $\Lambda$CDM parameters from \textit{Planck}-2018 measurements of the primary CMB alone (TT,TE,EE+lowE; no CMB lensing) \cite{Planck2018}. We obtain and marginalize this prior over $\tau_{re}$ using the code \texttt{FishLSS}\footnote{https://github.com/NoahSailer/FishLSS/} \cite{FishLSS}. Without including any further prior on $\mnus$, as given in Table \ref{table:paramserror-ksz+pl}, \Corr{we forecast $1\sigma$ constraints of $\sim 159$ meV and 129 meV on $\mnus$ for SO$\times$WISE and CMB-S4$\times$WISE, respectively} using the projected-fields kSZ bispectrum, again while restricting galaxy modes to the linear regime only. Unlike the corresponding $\mnus$ constraints from the kSZ$^2$ estimator (168 meV and 100 meV, respectively) \cite{P23}, those from $\BTTg$ here are immune to modeling uncertainties in HOD parameters and non-linearities in galaxy bias, which should otherwise be modified to accurately account for the effect of massive neutrinos (e.g.\,\cite{dvorkin}).

We also perform Fisher forecasts with the $\BTTg$ estimator by replacing the primary CMB prior on $\Lambda$CDM from \textit{Planck} with a forecasted one that combines low-$\ell$ and high-$\ell$ primary CMB information from LiteBIRD \cite{litebird} and SO/CMB-S4 respectively \cite{FishLSS}. From Table \ref{table:paramserror-ksz+pl}, we forecast \Corr{$1\sigma$ constraints of $\sim 97$ meV and 82 meV on $\mnus$ for SO$\times$WISE and CMB-S4$\times$WISE, respectively}. These are significantly tighter than those using the primary \textit{Planck} prior; this is expected given the high sensitivity of LiteBIRD at mapping $\ell<200$ modes and providing cosmic-variance limited estimates of $\tau_{re}$ \cite{litebird}, which improve the primary CMB prior on $A_s$. 

Finally, following a recent work on kSZ tomography \cite{Tishue25_mnu}, \Rev{we also consider an external prior combining information on $\Lambda$CDM+$\mnus$ parameters from primary \textit{Planck} data and upcoming Baryon Acoustic Oscillations (BAO) observations with DESI \cite{DESI2024_cosmo}; we obtain both of these priors using the package \texttt{pyfisher}\footnote{https://github.com/msyriac/pyfisher}. The forecasted 1$\sigma$ errors on $\mnus$ using this combined prior tighten from $\sim$75 meV [\textit{Planck} + DESI BAO] to about 54 meV when the kSZ information is added using the proposed $\BTTg$ estimator}\footnote{While the $\sigma(\mnus)$ constraint from the [\textit{Planck} + DESI BAO] prior only improves by 2\% once the $\BTTg$ signal is added, the combined constraint improves by around 28\% when a realistic Gaussian prior on the amplitude $A\propto\tau^2$ of $\BTTg$ is included.}. This would allow distinguishing between the normal/inverted neutrino hierarchies at a 1$\sigma$ level. 

Therefore, the $\BTTg$ estimator could significantly tighten constraints on neutrino masses by breaking parameter degeneracies in other competitive probes (see e.g.~\cite{mishra2018,mnu_CMBLSS,mnu_whitep}) such as BAO (as shown here), CMB lensing (e.g. ACT \cite{ACTDR6_lens_cosmomnu}) and shear measurements from LSS surveys (e.g.\,LSST \cite{VRO2019}). Thus, we demonstrate here that the projected-fields kSZ bispectrum can potentially become a powerful, complementary probe of neutrino masses through their impact on the linear growth rate and clustering of matter. 

\section{Discussion and Future Outlook}\label{discuss} 
In this work, we have proposed a novel kSZ (`full') bispectrum of the form $\langle TTg \rangle$ that cross-correlates two powers of a cleaned CMB temperature map (`$T$') with one field of a tracer of LSS density projected along the LOS (`g') in harmonic space. Since it does not require individual redshifts of objects, it is applicable to any LSS tracer with a known statistical redshift distribution. 

In Section \ref{theory}, we implemented our kSZ estimator using the binned bispectrum formalism, improving upon the existing projected-fields power spectrum ($\CTTg$) by avoiding its convolution over the CMB filter and its compression of information across different triangle shapes. Thus, our new $\BTTg$ estimator contains richer information that is better separated across different scales, given that it is a `full' bispectrum that goes beyond simply squaring and cross-correlating the CMB map. Therefore, we expect that scale-dependent cosmological signatures impacting the signal at intermediate scales (such as $\mnus$ above) are less likely to be biased by uncertainties in the baryonic distribution that dominate at small scales, and vice versa. Moreover, this bispectrum approach gives a better handle over some systematics; for example, choosing to restrict the galaxy modes to linear scales makes our results immune to modeling uncertainties in nonlinear bias and HOD, without compromising on the SNR. 

The $\BTTg$ estimator can provide complementary constraints on neutrino masses (as demonstrated in Section \ref{cosmo-sec}) and potentially on dark energy or modified gravity models via their impact on the linear growth rate. Although not explicitly shown here, the $\BTTg$ bispectrum could alternatively be used to measure the late-time baryon abundance ($f_b$) or baryon density profile in a robust way, improving upon the $\CTTg$ method \cite{Hill2016, Kusiak2021, Bolliet2022}. The baryon dependence could be modeled more accurately by incorporating the improved theoretical model (Appendix \ref{deriv}) for this estimator in a Halo Model framework using the code \texttt{class_sz}\footnote{\url{https://github.com/CLASS-SZ}}. 

Moreover, inferences from future measurements of the $\BTTg$ signal using MCMC analyses will require a further speed-up of our numerical implementation, possibly by utilizing the \texttt{cosmopower} emulation framework \cite{cosmopower, cosmopowerboris}. We leave these improvements to future work. \Rev{A necessary follow-up for interpreting the signal from future measurements with the proposed estimator would be to apply it to accurate simulated maps (e.g. Websky \cite{websky} or Agora \cite{agora}) and validate its pipeline. Certain key aspects to be quantitatively tested using simulations include estimating the kSZ bispectrum's total covariance matrix and the amount of non-Gaussian contribution to it due to CMB lensing, as well as quantifying the level of contamination due to residual foregrounds, particularly residual CIB, which may be present in post-ILC CMB maps and could bias the measured kSZ signal.}

Various photometric samples could be used in place of the WISE catalog for the LSS field, including those spanning higher redshifts such as the unWISE green or red subsamples \cite{unWISE2019}, DESI photometric LRGs \cite{DESILRG_2023}, or from the upcoming Rubin Observatory (VRO) \cite{VRO2019} or SPHEREx \cite{spherex2014} surveys, among others. \Rev{Such measurements should also account for a contribution due to the magnification bias of galaxies, which may become significant at higher redshifts ($z\gtrsim1$), as computed and suggested in \cite{Kusiak2021}}. A possible refinement would be to use a more optimal redshift weighting for these deeper surveys \cite{F16}, since the projected bispectrum can otherwise receive a majority contribution from $z<0.5$ or so \cite{buch2000, Dore2004, P23}. 

While galaxies are expected to yield the highest detection significance with this estimator, any LSS tracer of matter density can be used instead, such as galaxy/CMB weak lensing convergence (as shown for the kSZ$^{2}$ estimator in \cite{Bolliet2022, F16}) or quasars. In the future, the same framework could potentially be applied to 21-cm maps as the tracer, as suggested in the related work \cite{LaPlante_2020_21kSZkSZ_BiS}, although we note that it is necessary to use the improved model for this bispectrum (Appendix A and \cite{P23}) in such analyses to accurately interpret the kSZ signal contained in super-squeezed triangles from the epoch of reionization. \Rev{Moreover, in its current form, most of the relevant 21-cm-kSZ-kSZ bispectrum modes will be lost due to foreground wedge removal in 21-cm data; this issue can be avoided by considering the kSZ$^{2}\times\delta_g^{2}$ trispectrum instead, as suggested recently in \cite{ZhouLaPlante2025_kSZ221cm2}.} 

In conclusion, we have proposed a novel binned bispectrum to detect the late-time kSZ effect in cleaned CMB maps that uses a projected density field, thus making it applicable to a variety of LSS tracers. We forecasted high cumulative SNRs ($\sim$100-200) for the upcoming SO and CMB-S4 with this estimator, when combined with the WISE galaxy field restricted to linear scales. We showed that the SNR peaks for squeezed triangles in harmonic space that have a short (linear) LSS density mode and two long CMB temperature modes, \Corr{while the lensing correction is found to be small compared to the signal} for such high-SNR triangle configurations. We further illustrate the future potential of this kSZ bispectrum beyond measuring the baryonic abundance at cosmic times, by forecasting initial constraints on the sum of neutrino masses. Thus, the estimator proposed in this work contributes further to the ever-increasing scope of analyses employing the kSZ effect as a probe of baryonic astrophysics and beyond-$\Lambda$CDM cosmology.  

\begin{acknowledgments}
We are grateful to Simone Ferraro, Will Coulton, Mathew Madhavacheril, and the anonymous referee for their helpful feedback. We thank Boris Bolliet, Fiona McCarthy, and Aleksandra Kusiak for useful conversations. NB acknowledges support from NASA grants 80NSSC18K0695 and 80NSSC22K0410. JCH acknowledges support from NSF grant AST-2108536, NASA grant 80NSSC22K0721 (ATP), the Sloan Foundation, and the Simons Foundation.
\end{acknowledgments}
\appendix
\section{Improved Model of $\Blos$} \label{deriv}
The improved model of the $\Blos$ statistic that underlies our $\BTTg$ estimator was derived in the previous work \cite{P23}, which showed that it leads to significant ($\sim$10-15$\%$ level) scale-dependent changes in the predicted signal for the related $\CTTg$ estimator, in the context of upcoming measurements with SO and CMB-S4. 
Here, we summarize the improved model and some of its key aspects for completeness and refer the reader to \cite{P23} for a full derivation and discussion. 

Under the Limber approximation, the LOS bispectrum $\Blos = (1/2)\Bperp$ \cite{Hu_2000_ksz, Ma2002, D05, P23}, where $p_{\perp}$ is the transverse component of the electron momentum $\mathbf{p} \sim \delta\mathbf{v}$. Thus, the statistic is essentially a 5-point function of the form $\langle \delta\mathbf{v}\delta\mathbf{v}\delta \rangle$, whose (tree-level) Wick expansion consists of $\binom 52 = 10$ different terms \cite{D05}, out of which four are non-zero (see Table 1 in \cite{P23}). 

A rigorous derivation accounting for the directions of electron velocity fields leads to the following expression for the leading order (or `usual') term: 
\begin{align}\label{usual}
    & \Bperp^{\mathrm{usual}}(\mathbf{k_{1}}, \mathbf{k_{2}}, \mathbf{k_{3}}) = \nonumber \\
    & \int\frac{d^{3}k}{\left(2\pi\right)^{3}}\left[\sqrt{1-\mu_{1}^{2}}\sqrt{1-\mu_{2}^{2}} \right]P_{vv}(k) B_{\delta\delta\delta}(\mathbf{k_{1}}-\mathbf{k} \mathbf{k_{2}}+\mathbf{k},\mathbf{k_{3}}),
\end{align}
where $\mu_{\alpha} \equiv \hatbf{k}_{\alpha} \cdot \hatbf{k}$ for $\alpha$= 1, 2. The other three non-zero terms are referred to as the `extra$-i$' terms, because they each have a rough geometric scaling {$\propto 1/k_i$} for $i = 1, 2, 3$. Two of these are symmetric in $(\mathbf{k_{1}}, \mathbf{k_{2}})$ and are given by:
\begin{dmath} \label{extra-1}
  \Bperp^{\mathrm{extra-1}}(\mathbf{k_{1}}, \mathbf{k_{2}}, \mathbf{k_{3}}) \\
  = \int \frac{d^{3}k'}{\left( 2\pi \right) ^{3}} \left[\sqrt{1 - \mu_{1}^{2}}\sqrt{1 - \mu_{2}^{2}} \frac{k'}{|\mathbf{k'}+\mathbf{k}_{1}|}\right] \\
  P_{\delta v}(k') 
  B_{v\delta\delta }( \mathbf{k_{1}}+\mathbf{k'}, \mathbf{k_{2}}-\mathbf{k'},\mathbf{k_{3}}),
\end{dmath}
where $\mu_{\alpha} \equiv \hatbf{k}_{\alpha} \cdot \hatbf{k'}$ for $\alpha$ = 1, 2, and 
\begin{dmath} \label{extra-2}
  \Bperp^{\mathrm{extra-2}}(\mathbf{k_{1}}, \mathbf{k_{2}}, \mathbf{k_{3}}) \\
  = \int \frac{d^{3}k}{\left( 2\pi \right) ^{3}} \left[\sqrt{1 - \mu_{1}^{2}}\sqrt{1 - \mu_{2}^{2}} \frac{k}{|\mathbf{k}+\mathbf{k}_{2}|}\right] \\
  P_{v\delta}(k) 
  B_{\delta v \delta }( \mathbf{k_{1}}-\mathbf{k}, \mathbf{k_{2}}+\mathbf{k},\mathbf{k_{3}}),
\end{dmath}
where $\mu_{\alpha} \equiv \hatbf{k}_{\alpha} \cdot \hatbf{k}$ for $\alpha$ = 1, 2. The extra$-3$ term has a slightly different form since $\mathbf{k_{3}}$ corresponds to $\delta_{g}$ and not the kSZ modes, and can be positive or negative:
\begin{align}\label{extra-3}
    & \Bperp^{\mathrm{extra-3}}(\mathbf{k_{1}}, \mathbf{k_{2}}, \mathbf{k_{3}}) \nonumber \\
    & = \int \frac{d^{3}k}{\left( 2\pi \right) ^{3}} 
  \bigg[- \sqrt{1 - \mu_{1}^{2}}\sqrt{1 - \mu_{2}^{2}} \frac{k}{|\mathbf{k}+\mathbf{k}_{3}|} + \nonumber\\ 
  & ~~~~~~~~~~~~~~~~~~ \sqrt{1 - \mu_{1}^{2}}\sqrt{1 - \mu_{12}^{2}}\frac{k_{1}}{|\mathbf{k}+\mathbf{k}_{3}|}\bigg] 
  \nonumber \\
  & ~~~~~~~~~  P_{\delta\delta}(|k-k_{1}|)
  B_{vv \delta }( \mathbf{k}, \mathbf{k_{1}}+\mathbf{k_{2}}-\mathbf{k},\mathbf{k_{3}}),
\end{align}
where $\mu_{\alpha} \equiv \hatbf{k}_{\alpha} \cdot \hatbf{k}$ for $\alpha$ = 1, 2, and $\mu_{12} \equiv \hatbf{k}_{1} \cdot \hatbf{k}_{2}$.

Overall, the improved model predicts that the $\Blos$ statistic is given by:
\begin{equation} \label{full}
\Blos = (1/2) \left[\Bperp^{\mathrm{usual}}+ \Bperp^{\mathrm{extra-1}} + \Bperp^{\mathrm{extra-2}} + \Bperp^{\mathrm{extra-3}} \right].
\end{equation}

\section{Analytical Binned Bispectrum in the Flat-sky Limit}\label{binflat}
We analytically defined the binning of our projected-fields $\BTTg$ bispectrum in the flat-sky limit in Section \ref{ourest}. Here, we show that this definition (Eq.\eqref{binning}) generalized is consistent with previous works following the binned bispectrum approach \cite{bucher_2010, Bucher_2016, Coulton, CoultonSpergel2019}.

In general, any 2D field across the sky (e.g. primary CMB maps, $\Theta_f$, $\delta_g$) can be decomposed using spherical harmonics ($Y_{\ell m}(\hatbf{n})$). 
The all-sky bispectrum between any such maps $(X, Y, Z)$ is then defined as the ensemble average of the products of their spherical harmonic coefficients and can be written (e.g.\cite{Bucher_2016, CoultonSpergel2019}) using the Gaunt integral\footnote{The Gaunt integral is expressed using Wigner-3j symbols as:
\begin{align}
\mathcal{G}^{m_1,m_2,m_3}_{\ell_1,\ell_2,\ell_3}& \equiv \int\mathrm{d}^2\mathbf{n}Y_{\ell_1m_1}(\mathbf{n}) Y_{\ell_2m_2}(\mathbf{n})  Y_{\ell_3m_3}(\mathbf{n})\nonumber\\ &= \sqrt{\frac{(2\ell_1+1)(2\ell_2+1)(2\ell_3+1)}{4\pi}} \nonumber\\ & \;\;\;\;\;\;\; \begin{pmatrix}
    \ell_1&\ell_2&\ell_3 \\
    m_1&m_2& m_3
  \end{pmatrix}\begin{pmatrix}
    \ell_1 & \ell_2 & \ell_3 \\
    0 & 0 & 0
  \end{pmatrix} \nonumber\\ &= \sqrt{N_{\Delta}^{\ell_1\ell_2\ell_3}}\begin{pmatrix}
    \ell_1&\ell_2&\ell_3 \\
    m_1&m_2&m_3
  \end{pmatrix}. 
\end{align}} 
as:
\begin{equation}\label{allsky}
    \mathscr{B}_{\ell_1\ell_2\ell_3}^{X, Y, Z} = \sqrt{N_{\Delta}^{\ell_1\ell_2\ell_3}}\sum_{m}\begin{pmatrix}
    \ell_1 & \ell_2 & \ell_3 \\
    m_1 & m_2 & m_3
  \end{pmatrix} \langle a^{X}_{\ell_1 m_1}a^{Y}_{\ell_2 m_2}a^{Z}_{\ell_3 m_3} \rangle.
\end{equation}
(The all-sky bispectrum can also be defined without the $\sqrt{N_{\Delta}}$ factor \cite{hu_2000_bis}, but we follow the above convention throughout). 

On the other hand, the flat-sky or `reduced' bispectrum is defined as in Eq.\eqref{diracbi} and is given in general by:
\begin{equation}\label{defn_alms}
B_{\ell_1\ell_2\ell_3}^{X, Y, Z} = \frac{\sum_{m}\begin{pmatrix}
    \ell_1 & \ell_2 & \ell_3 \\
    m_1 & m_2 & m_3
  \end{pmatrix} \langle a^{X}_{\ell_1 m_1}a^{Y}_{\ell_2 m_2}a^{Z}_{\ell_3 m_3} \rangle}{\sqrt{N_{\Delta}^{\ell_1\ell_2\ell_3}}}
\end{equation}
so that for a single multipole triplet, the flat-sky and all-sky bispectra are related \cite{hu_2000_bis, Bucher_2016}: 
$B_{\ell_1\ell_2\ell_3}^{X, Y, Z} = \left(\mathscr{B}_{\ell_1\ell_2\ell_3}^{X, Y, Z}/N_{\Delta}^{\ell_1\ell_2\ell_3}\right)$. 

Now, by binning the harmonic space in 3D similar to Section \ref{ourest}, the all-sky binned bispectrum $\mathscr{B}_{abc}^{X, Y, Z}$ is effectively an average over all possible $\mathscr{B}_{\ell_1\ell_2\ell_3}^{X, Y, Z}$ that lie within the 3D $\ell$-bin \cite{bucher_2010, Bucher_2016}. Substituting the relation between flat-sky and all-sky bispectra for each multipole triplet, we recover our definition for a binned flat-sky bispectrum (Eq.\eqref{binning}) for fields $(X, Y, Z)$ in general:
\begin{equation}\label{binning-gen}
B^{X, Y, Z}_{abc} = \frac{1}{N_{abc}} \left(\sum_{\substack{\ell_{1}\in\Delta_{a}\\\ell_{2}\in\Delta_{b}\\\ell_{3}\in\Delta_{c}}} N_{\Delta}^{\ell_1\ell_2\ell_3} B_{\ell_1\ell_2\ell_3}^{X, Y, Z} \right). 
\end{equation}
It is thus consistent with the all-sky formalism \cite{bucher_2010, Bucher_2016}, and is a \textit{weighted} average of all $B_{\ell_1\ell_2\ell_3}^{X, Y, Z}$ within the bin. 

The binned approach has also been extended to apply directly to (simulated or real data) maps in the flat-sky limit \cite{Coulton, CoultonSpergel2019}, which is implemented by `filtering' the maps to select only those $\ell$ modes that lie within the bin: $W^Z_{c}(\mathbf{n}) = \sum\limits_{\ell\in\Delta_{c}} \sum\limits_{m}{Y}_{\ell,m}(\mathbf{n})a^Z_{\ell,m}$. The flat-sky bispectrum is then estimated as:
\begin{equation}
\hat{B}^{X, Y, Z}_{abc} = \int\mathrm{d}^2\mathbf{n}\frac{1}{N_{abc}}W^X_{a}W^Y_{b}W^Z_{c},
\end{equation}
which is seen to be equivalent to our analytical definition Eq.\eqref{binning-gen} by using the Gaunt integral. 

\section{CMB lensing contribution to the Projected-Fields Bispectrum and its Covariance Matrix}\label{lens}
As mentioned in Section \ref{lens_main}, when the projected-fields kSZ bispectrum is estimated from observed CMB maps, there will potentially be an additional contribution to the signal due to the weak lensing of the CMB by the intervening LSS. To compute this lensing correction to our estimator, we follow \cite{F16} and analogously expand up to first order in the lensing potential $\psi$ \cite{Lewis2006}: $\Theta(\mathbf{x}) = \tilde{\Theta}(\mathbf{x}) + \nabla\psi\cdot\nabla\tilde{\Theta}(\mathbf{x})+ ...$, where the unlensed and lensed CMB anisotropies are again denoted by $\tilde{\Theta}$ and $\Theta$ respectively. In Fourier space, 
\begin{equation}\label{exp}
[\nabla\psi\cdot\nabla\tilde{\Theta}](\mathbf{L}) = - \int\frac{\mathrm{d}^2\mathbf{L'}}{(2\pi){}^2} \mathbf{L'}\cdot(\mathbf{L}-\mathbf{L'})\psi(\mathbf{L'})\tilde{\Theta}(\mathbf{L}-\mathbf{L'}).
\end{equation}
Now, consider the lensed signal, from Eq.\eqref{diracbi}, 
\begin{equation}
\langle \Theta_b(\mathbf{\ell_1})\,\Theta_b(\mathbf{\ell_2})\,\delta_g(\mathbf{\ell_3})\rangle = b(\ell_1)b(\ell_2)\langle \Theta(\mathbf{\ell_1})\,\Theta(\mathbf{\ell_2})\,\delta_g(\mathbf{\ell_3})\rangle.
\nonumber
\end{equation}
Expanding up to first order in $\psi$, 
\begin{align}\label{lens_exp}
\langle \Theta(\mathbf{\ell_1})\,\Theta(\mathbf{\ell_2})\,\delta_g(\mathbf{\ell_3})\rangle = \,\,\,\,\,\,\,\,\,\,\,\{\langle \tilde{\Theta}(\mathbf{\ell_1})\,\tilde{\Theta}(\mathbf{\ell_2})\,\delta_g(\mathbf{\ell_3})\rangle \nonumber\\ +\,\langle[\nabla\psi\cdot\nabla\tilde{\Theta}](\mathbf{\ell_1})\tilde{\Theta}(\mathbf{\ell_2})\,\delta_g(\mathbf{\ell_3})\rangle \nonumber\\ +\,(\mathbf{\ell_1}\leftrightarrow\mathbf{\ell_2}) \}.
\end{align}
The first term on the right-hand-side sources the fiducial (unlensed) projected-fields kSZ bispectrum that we computed in Eq.\eqref{bis}. The third term is symmetric to the second term with the roles of $\mathbf{\ell_1}$ and $\mathbf{\ell_2}$ interchanged; together, these two additional terms source the lensing correction to our estimator. Substituting Eq.\eqref{exp},
\begin{dmath}\label{simplify}
\langle[\nabla\psi\cdot\nabla\tilde{\Theta}](\mathbf{\ell_1})\tilde{\Theta}(\mathbf{\ell_2})\,\delta_g(\mathbf{\ell_3})\rangle = - \int\frac{\mathrm{d}^2\mathbf{L'}}{(2\pi){}^2} \mathbf{L'}\cdot(\mathbf{\ell_1}-\mathbf{L'})\langle\psi(\mathbf{L'})\tilde{\Theta}(\mathbf{\ell_1}-\mathbf{L'})\tilde{\Theta}(\mathbf{\ell_2})\,\delta_g(\mathbf{\ell_3})\rangle.   
\end{dmath}

\begin{figure}
\center
{
\includegraphics[width=0.98\columnwidth]{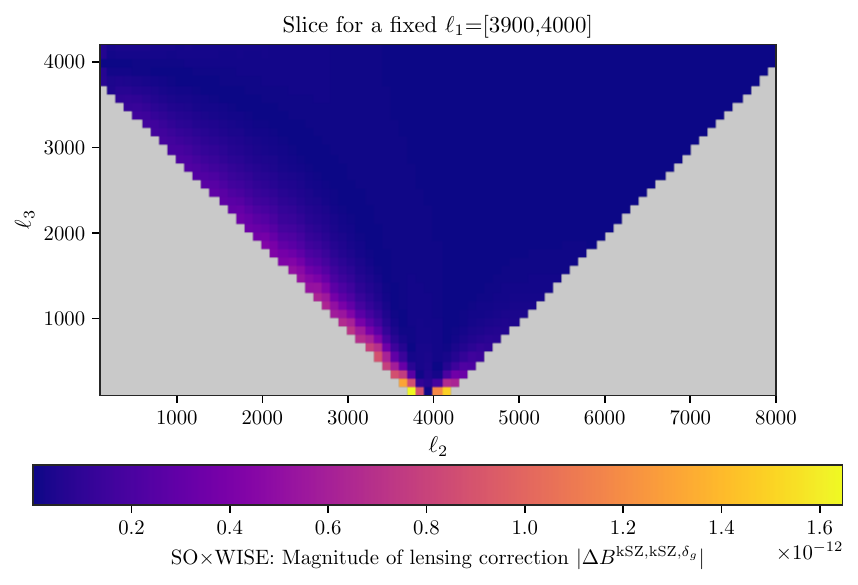} 
}
\caption{A 2D slice in harmonic space in the extended scenario with a fixed $\ell_1$-bin of [3900,4000]. \Corr{The absolute value of the lensing contribution in each bin}, $|\Delta\BTTg|$, (in units of $\mu$K$^2$) is visualized here for SO$\times$WISE; it peaks for squeezed triangle configurations having both CMB modes $\lesssim 4500$.} 
\label{fig:lens_magnitude}
\end{figure}

\begin{figure*}
    \center
    {
    \includegraphics[width=0.64\textwidth]{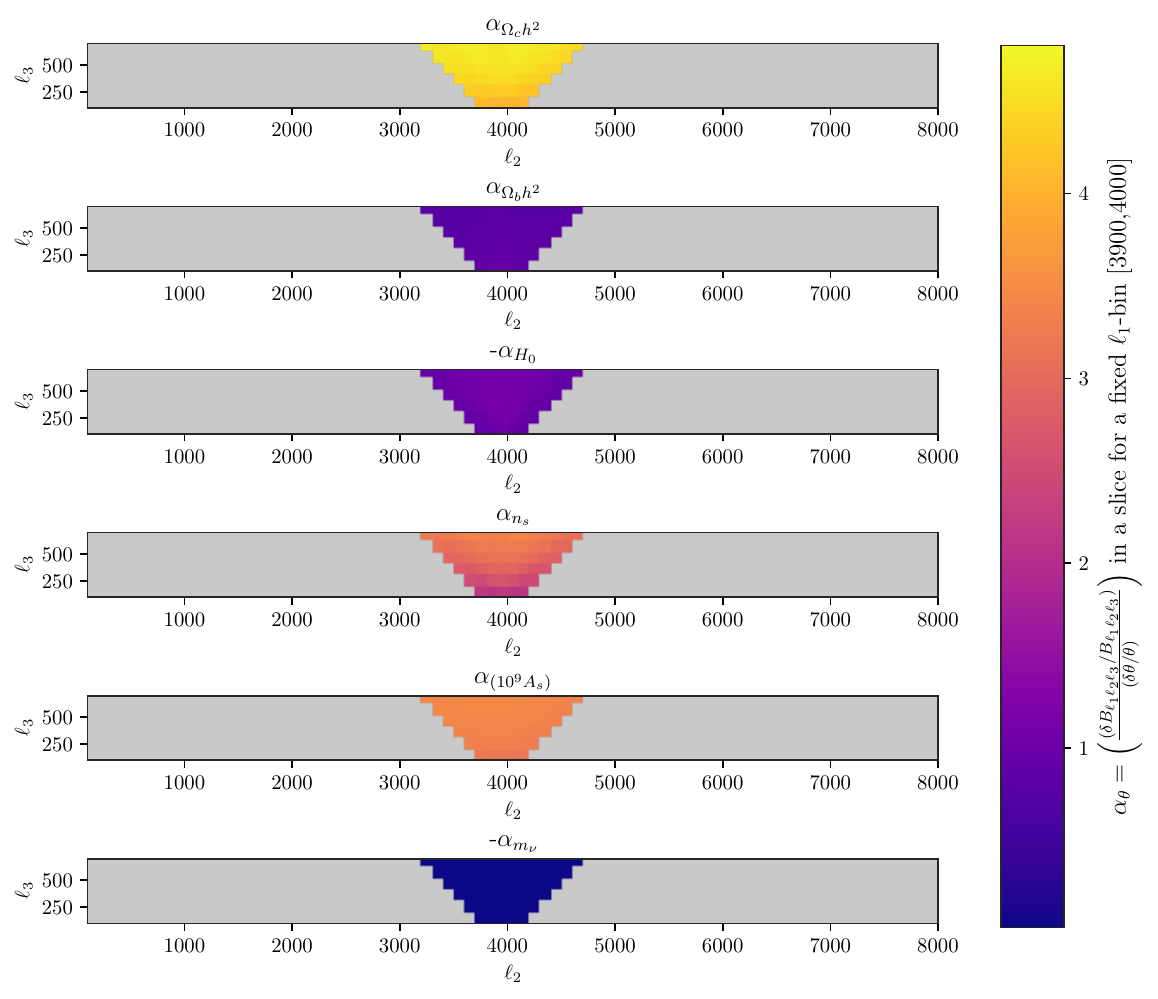}
    }
    \caption{Power-law indices $\alpha_{\theta}$ for each of the $\Lambda$CDM parameters $\pmb{\theta} = \{H_{0}, \Omega_{b}h^{2}, \Omega_{c}h^{2}, 10^{9}A_{s}, n_{s}\}$ and $\mnus$, showing the approximate dependence of $\BTTg$ on them in a fixed-$\ell_1$ slice (that spans the entire default range of $\ell_2$ and $\ell_3$) for SO$\times$WISE. The indices are negative for $H_{0}$ and $\mnus$.}
    \label{fig:alphas}
\end{figure*}

\begin{figure*} 
  \centering
  \subfloat[Fixed Cosmology]{\includegraphics[width=0.40\textwidth]{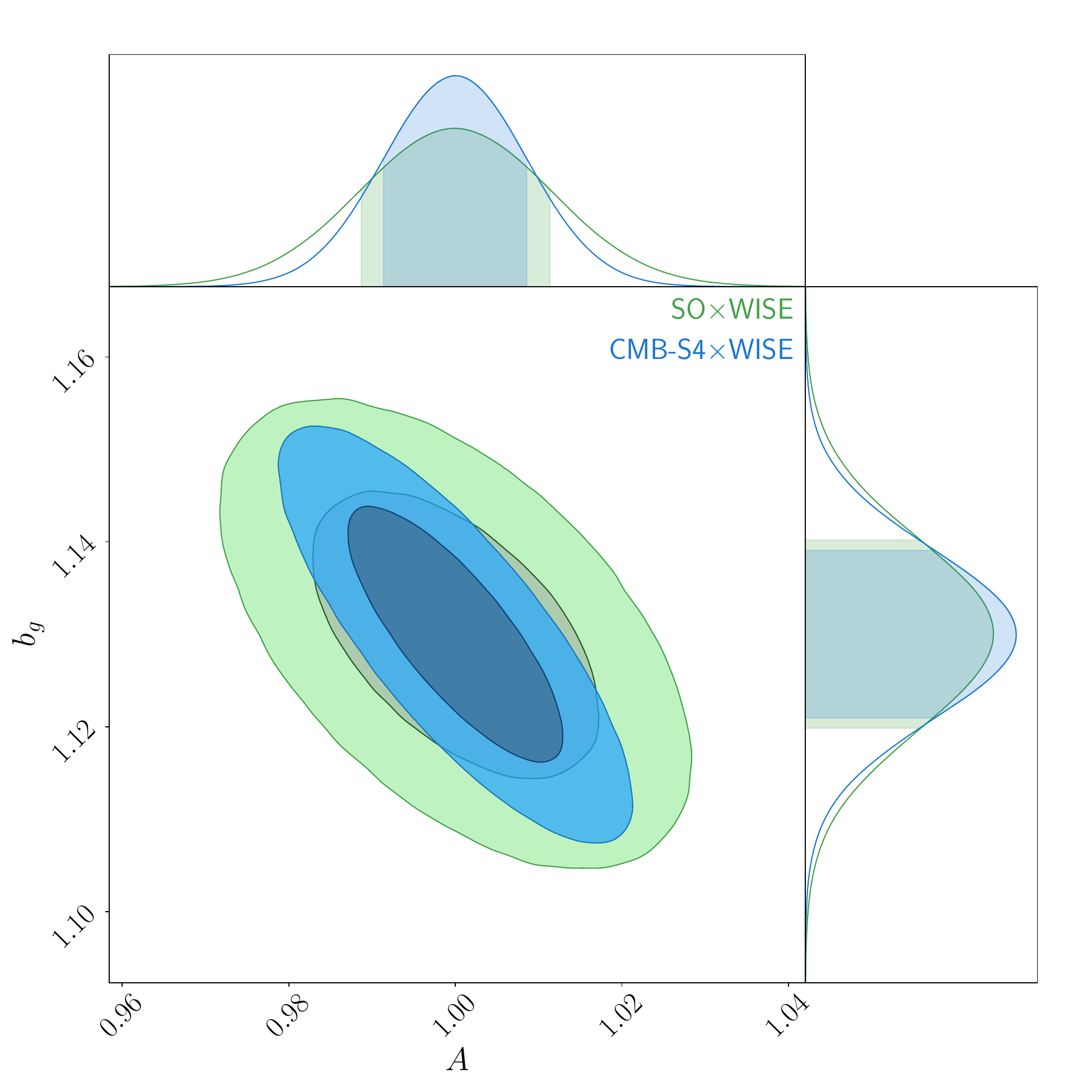}\label{fig:f1}}
  \subfloat[Marginalizing over $\Lambda$CDM Parameters]{\includegraphics[width=0.40\textwidth]{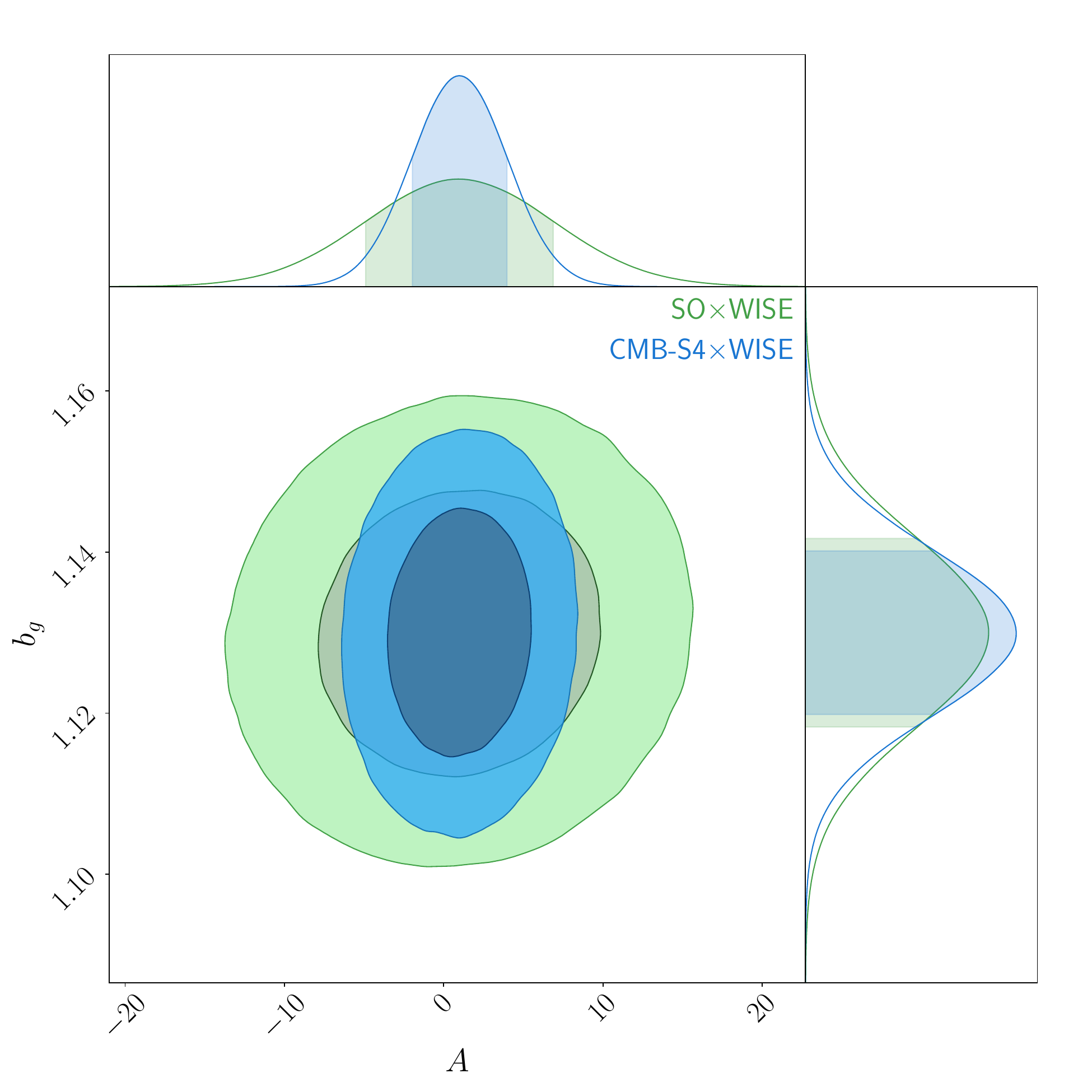}\label{fig:f2}}
  \caption{\Corr{Contours for SO$\times$WISE (green) and CMB-S4$\times$WISE (blue) show forecasted 68$\%$ (darker) and 95$\%$ confidence levels for $\{A, b_g\}$ assuming a fixed cosmology (left), and by modeling the cosmological dependence of the $\BTTg$ signal and marginalizing over $\Lambda$CDM parameters (right), without applying any external cosmological prior here.} Fisher analysis is done in the default regime, and assuming realistic post-ILC noise for SO and CMB-S4.} 
   \label{fig:A-bg}
\end{figure*}

By Wick's theorem, the four-point correlation of the form $\langle\psi\tilde{\Theta}\tilde{\Theta}\delta_g\rangle$ appearing above can be decomposed \cite{F16} into a non-zero contraction of the form $\langle\psi\delta_g\rangle\langle\tilde{\Theta}\tilde{\Theta}\rangle$, along with other terms that do not strictly vanish but are negligible at our scales of interest where the ISW effect is sub-dominant: contractions of the form $\langle\psi\tilde{\Theta}\rangle\langle\tilde{\Theta}\delta_g\rangle$ and a connected four-point function. For the dominant term contributing to $\langle\psi\tilde{\Theta}\tilde{\Theta}\delta_g\rangle$, we have:
\begin{equation}
 \langle\psi(\mathbf{L'})\delta_g(\mathbf{\ell_3})\rangle = (2\pi){}^2 C^{\psi\delta_g}_{\ell_3}\delta_{D}(\mathbf{L'}+\mathbf{\ell_3}), \nonumber
\end{equation}
\begin{equation}
\langle\tilde{\Theta}(\mathbf{\ell_1}-\mathbf{L'})\tilde{\Theta}(\mathbf{\ell_2})\rangle = (2\pi){}^2 C_{\ell_2}^{\tilde{T}\tilde{T}} \delta_{D}(\mathbf{\ell_1}-\mathbf{L'}+\mathbf{\ell_2}). \nonumber
\end{equation}
Substituting these back into Eq.\eqref{simplify} and recalling the Dirac delta function in Eq.\eqref{diracbi}, we thus calculate the leading order correction due to lensing from Eq.\eqref{lens_exp} to be:
\begin{equation}
\Delta\BTTg_{\ell_1\ell_2\ell_3} = -b(\ell_1)b(\ell_2)C_{\ell_3}^{\psi\delta_g}\left( \mathbf{\ell_3}\cdot\mathbf{\ell_1}C_{\ell_1}^{\tilde{T}\tilde{T}}+\mathbf{\ell_3}\cdot\mathbf{\ell_2}C_{\ell_2}^{\tilde{T}\tilde{T}} \right) \nonumber
\end{equation}
where we have also included the other symmetric term's contribution $(\mathbf{\ell_1}\leftrightarrow\mathbf{\ell_2})$, and simplified further using the closed triangle condition $(\ell_1+\ell_2+\ell_3 = 0)$. For any multipole-triplet $(\ell_1, \ell_2, \ell_3)$, the lensing term is computed by using CAMB for the cosmological power spectra, and by using the cosine rule: $\mathbf{\ell_3}\cdot\mathbf{\ell_1} = \frac{(\ell_2^{2}-\ell_1^{2}-\ell_3^{2})}{2}$ and $\mathbf{\ell_3}\cdot\mathbf{\ell_2} = \frac{(\ell_1^{2}-\ell_2^{2}-\ell_3^{2})}{2}$. 

Since the lensing correction can be positive or negative, we plot the absolute value of the binned $(|\Delta\BTTg|)$ for 3D bins in a 2D slice with a fixed $\ell_1$ = [3900,4000] in Fig.\,\ref{fig:lens_magnitude}. We see that the lensing contribution is higher in magnitude for squeezed triangles as compared to other shapes. As discussed further in Section \ref{fore}, we find that the lensing correction is smaller than the projected-fields bispectrum signal for all bins that contain a significant kSZ signal.

We now consider the lensing contribution to the estimator's covariance, which introduces a non-Gaussianity of the form $\langle a^{\Theta_b}_{\ell_1 m_1}a^{\Theta_b}_{\ell_2 m_2}a^{\Theta_b *}_{\ell_1' m_1'}a^{\Theta_b *}_{\ell_2' m_2'}\rangle \langle a^{\delta_g}_{\ell_3 m_3}a^{\delta_g *}_{\ell_3' m_3'} \rangle$, as given in Section \ref{variance}. We refer the reader to \cite{Coulton+nGCovar2020} (Appendix A.1 therein) for details of their calculation of this term in their separate context of primary CMB $\langle TTT\rangle$ bispectra. We follow their derivation for the connected temperature trispectrum $\langle T(\ell_1)T(\ell_2)T^{*}(\ell'_1)T^{*}(\ell'_2) \rangle_{c}$ (see Eq.(A6) therein), and similarly drop its sub-dominant 1-loop corrections. We then derive the leading-order lensing contribution in our case to be:
\begin{align}\label{lenscorr-var}
   & = (2\pi)^{8}\delta_{D}(0)\delta_{D}(\mathbf{\ell_3}-\mathbf{\ell'_3})\delta_{D}\left(\sum_{i=1}^{3} \mathbf{\ell_i}\right)\delta_{D}\left(\sum_{i=1}^{3} \mathbf{\ell'_i}\right) \nonumber \\
   & C_{\ell_3}^{\delta_g\delta_g}C_{\ell_3}^{\psi\psi} \mathbf{\ell_3}\cdot\mathbf{\ell_1}C_{\ell_1}^{\tilde{T}\tilde{T}}  
   \mathbf{\ell_3}\cdot\mathbf{\ell'_1}C_{\ell'_1}^{\tilde{T}\tilde{T}}
   + 3 \,\, \mathrm{permutations},
\end{align}
which is analogous to the leading order term in Eq.(A9) of \cite{Coulton+nGCovar2020}. Note that the number of possible permutations (from symmetry considerations) in the case of our $\BTTg$ estimator is just 4, instead of 36 for the CMB temperature bispectrum \cite{Coulton+nGCovar2020}; this is because in our case, we only have 2 lensed temperature fields instead of 3, which reduces the average amplitude (and multiplicity) of this lensing term by an overall factor of 9. The expression above has a mathematically similar form to that of $\Delta\BTTg$. So, we intuitively expect that whenever the lensing contribution to the signal is small (e.g. for the high-SNR squeezed triangle types in Fig.\,1), the lensing-induced covariance term is also relatively small.  
\section{Understanding parameter dependency} \label{cosmodep}
To gain some basic intuition, we study the dependence of only the amplitude of the projected-fields kSZ bispectrum on $\Lambda$CDM parameters and $\mnus$. As mentioned in Section \ref{cosmo-sec}, for our purposes here, we model it simplistically as a power-law for each of these six parameters: $\BTTg_{\ell_1\ell_2\ell_3}\propto\theta^{\alpha_{\theta}}$. The power-law indices $\alpha_{\theta}$ are functions of $(\ell_1, \ell_2, \ell_3)$, and are numerically estimated as $\alpha_{\theta}\approx (\delta B/B)/(\delta\theta/\theta)$, where we have suppressed multipole and field names for clarity. We vary the $\Lambda$CDM parameters about their fiducial values using the same step sizes $\delta\theta$ as \cite{Madhavacheril2017, P23} and \cite{cromer_table} (Table I therein). $\delta B$ is the corresponding shift in the signal after one of these parameters is modified. 

Fig.\,\ref{fig:alphas} shows the power-law indices for different bins in a fixed-$\ell_1$ slice for SO$\times$WISE. Since the related $\CTTg$ estimator is a compressed version of the $\BTTg$ bispectrum, their overall amplitudes have power-law indices that are close to each other (comparing with Fig.\,6 in \cite{P23}). 
 Because $P_{\delta\delta}\propto A_s$, from Eq.\eqref{gil}, the bispectrum's amplitude approximately scales as $A_s^3$ or so, as seen here. It has the strongest dependence on $\Omega_{c}h^2$, with a high $\alpha_{\Omega_{c}h^2}\sim4.5$ value that varies slightly across the $\ell$-space; this is because of the non-linear dependence of the matter power spectrum as well as the linear growth rate $f$ on $\Omega_m = \Omega_{c}+\Omega_{b}$, where the density of cold dark matter contributes a factor of around 5 times that of baryons in the fiducial $\Lambda$CDM model. $\alpha_{n_s}$ varies the most within the $\ell$-space, since it is the slope of the primordial spectrum. $\mnus$ also imprints a scale dependence in the signal via $f$, but this effect is small compared to the fiducial $\Lambda$CDM parameters, which necessitates an external prior on $\Lambda$CDM (such as from \textit{Planck} primary CMB) for constraining $\mnus$. 

Now, we consider the case where a fixed cosmology is assumed, in order to simply constrain the baryonic abundance using the $\BTTg$ estimator's amplitude $A$. As explained in Sec.\,\ref{cosmo-sec}, we perform a Fisher analysis for $A$ and $b_g$ to forecast joint constraints at the level of $\sim1\%$ on both parameters (Fig \ref{fig:A-bg}, left). Next, we consider the case where the cosmological dependence is also included. We find that opening up the $\Lambda$CDM parameter space severely degrades the constraints on $A$ (Fig \ref{fig:A-bg}, right), partly because it is considerably degenerate with $\Omega_bh^2$; $b_g$ remains well-constrained due to the lensing correction. As discussed further in Sec.\,\ref{cosmo-sec}, we find that the combination $\tilde{A}\equiv (A\,b_g)$ is also well-constrained when jointly fitted with $\Lambda$CDM parameters, since $\BTTg \propto (A\,b_g)$ and $\tilde{A}$ acts as the total (overall) amplitude of the $\BTTg$ kSZ signal.

\bibliography{library}
\end{document}